\documentclass[superscriptaddress,aps,preprintnumbers,amsmath,amssymb,prd,nofootinbib,preprint]{revtex4-1}
\pdfoutput=1
\usepackage{graphicx}
\usepackage{epstopdf}
\usepackage{dcolumn}% Align table columns on decimal point
\usepackage{bm}% bold math
\usepackage{bbm}
\usepackage{hyperref}
\usepackage[usenames, dvipsnames]{color}
\usepackage{amsmath,amssymb}
\usepackage[sort&compress]{natbib}
\usepackage{float}
\usepackage{multirow}

\numberwithin{equation}{section}

\renewcommand{\thesubsection}{\thesection.\arabic{subsection}}

\makeatletter
\renewcommand{\p@subsection}{}
\renewcommand{\p@subsubsection}{}
\makeatother

\hypersetup{colorlinks,citecolor= blue,linkcolor= black}

\begin{document}

%%%%%%%%%%%%%%%%%%%%%%%%%%%%%%%%%%%%%%%%%%%

\def\a{\alpha}
\def\b{\beta}
\def\c{\varepsilon}
\def\d{\delta}
\def\e{\epsilon}
\def\f{\phi}
\def\g{\gamma}
\def\h{\theta}
\def\k{\kappa}
\def\l{\lambda}
\def\m{\mu}
\def\n{\nu}
\def\p{\psi}
\def\q{\partial}
\def\r{\rho}
\def\s{\sigma}
\def\t{\tau}
\def\u{\upsilon}
\def\v{\varphi}
\def\w{\omega}
\def\x{\xi}
\def\y{\eta}
\def\z{\zeta}
\def\D{\Delta}
\def\G{\Gamma}
\def\H{\Theta}
\def\L{\Lambda}
\def\F{\Phi}
\def\P{\Psi}
\def\S{\Sigma}

\def\o{\over}
\def\beq{\begin{align}}
\def\eeq{\end{align}}
\newcommand{\gsim}{ \mathop{}_{\textstyle \sim}^{\textstyle >} }
\newcommand{\lsim}{ \mathop{}_{\textstyle \sim}^{\textstyle <} }
\newcommand{\vev}[1]{ \left\langle {#1} \right\rangle }
\newcommand{\bra}[1]{ \langle {#1} | }
\newcommand{\ket}[1]{ | {#1} \rangle }
\newcommand{\EV}{ {\rm eV} }
\newcommand{\KEV}{ {\rm keV} }
\newcommand{\MEV}{ {\rm MeV} }
\newcommand{\GEV}{ {\rm GeV} }
\newcommand{\TEV}{ {\rm TeV} }
\newcommand{\1}{\mbox{1}\hspace{-0.25em}\mbox{l}}
\newcommand{\headline}[1]{\noindent{\bf #1}}
\def\diag{\mathop{\rm diag}\nolimits}
\def\Spin{\mathop{\rm Spin}}
\def\SO{\mathop{\rm SO}}
\def\O{\mathop{\rm O}}
\def\SU{\mathop{\rm SU}}
\def\U{\mathop{\rm U}}
\def\Sp{\mathop{\rm Sp}}
\def\SL{\mathop{\rm SL}}
\def\tr{\mathop{\rm tr}}
\def\mpl{M_{\rm Pl}}

\def\IJMP{Int.~J.~Mod.~Phys. }
\def\MPL{Mod.~Phys.~Lett. }
\def\NP{Nucl.~Phys. }
\def\PL{Phys.~Lett. }
\def\PR{Phys.~Rev. }
\def\PRL{Phys.~Rev.~Lett. }
\def\PTP{Prog.~Theor.~Phys. }
\def\ZP{Z.~Phys. }

\def\dd{\mathrm{d}}
\def\ff{\mathrm{f}}
\def\BH{{\rm BH}}
\def\inf{{\rm inf}}
\def\ev{{\rm evap}}
\def\eq{{\rm eq}}
\def\SM{{\rm sm}}
\def\Mpl{M_{\rm Pl}}
\def\GeV{{\rm GeV}}
\newcommand{\Red}[1]{\textcolor{red}{#1}}
\newcommand{\Blue}[1]{\textcolor{blue}{#1}}
\newcommand{\RC}[1]{\textcolor{blue}{\bf RC: #1}}
\newcommand{\MBA}[1]{{{\bf \color{ForestGreen} [MBA: #1]}}}
\newcommand{\mba}[1]{{\color{ForestGreen} #1}}

\newcommand{\dec}{{\rm dec}}
\newcommand{\nr}{{\rm nr}}

\newcommand{\nn}{\nonumber}
\newcommand{\bea}{\begin{eqnarray}}
\newcommand{\eea}{\end{eqnarray}}

\newcommand{\LC}{\ensuremath{\Lambda\mathrm{CDM}}}

\newcommand{\tot}{\mathrm{tot}}
\newcommand{\mpc}{\mathrm{Mpc}}

\newcommand{\dm}{\mathrm{dm}}
\newcommand{\cdm}{\mathrm{cdm}}
\newcommand{\wdm}{\mathrm{wdm}}
\newcommand{\DN}{\Delta N_\mathrm{eff}}

\newcommand{\ta}{{\tilde{a}}}
\newcommand{\grav}{{3/2}}

\newcommand{\mH}{\mathcal{H}}
\newcommand{\mO}{\ensuremath{\mathcal{O}}}
\newcommand{\mL}{\ensuremath{\mathcal{L}}}
\newcommand{\mV}{\ensuremath{\mathcal{V}}}
\newcommand{\mM}{\ensuremath{\mathcal{M}}}

\newcommand{\Sec}[1]{Sec.~\ref{#1}}
\newcommand{\Secs}[2]{Secs.~\ref{#1} and \ref{#2}}
\newcommand{\App}[1]{Appendix~\ref{#1}}
\newcommand{\Tab}[1]{Table~\ref{#1}}
\newcommand{\Fig}[1]{Fig.~\ref{#1}}
\newcommand{\Eq}[1]{Eq.~(\ref{#1})}
\newcommand{\Eqs}[2]{Eqs.~(\ref{#1}) and (\ref{#2})}
\newcommand{\Eqst}[2]{Eqs.~(\ref{#1})-(\ref{#2})}
\newcommand{\eg}{\textit{e.g.}\ }
\newcommand{\ie}{\textit{i.e.}\ }
\newcommand{\draftnote}[1]{\textbf{#1}}

\newcommand{\bl}{\left}
\newcommand{\br}{\right}

\title{
Common Origin of Warm Dark Matter and Dark Radiation
}
\preprint{LCTP-19-31}

\author{Manuel A.~Buen-Abad}
\email{manuel\_buen-abad@brown.edu}
\affiliation{Department of Physics, Brown University, Providence, RI, 02912, USA}
\author{Raymond T.~Co}
\email{rtco@umich.edu}
\affiliation{Leinweber Center for Theoretical Physics, University of Michigan, Ann Arbor, MI 48109, USA}
\author{Keisuke Harigaya}
\email{keisukeharigaya@ias.edu}
\affiliation{School of Natural Sciences, Institute for Advanced Study, Princeton, NJ 08540, USA}

\begin{abstract}
We consider a cosmological scenario where a relativistic particle and a stable massive particle are simultaneously produced from the decay of a late-decaying particle after Big-Bang Nucleosynthesis but before matter-radiation equality. The relativistic and massive particles behave as dark radiation and warm dark matter, respectively. Due to a common origin, the warmness and abundances are closely related. We refer to the models that lead to such a scenario as Common Origin of Warm and Relativistic Decay Products (COWaRD). We show that COWaRD predicts a correlation between the amount of dark radiation and suppression of the large scale structure, which can be tested in future precision cosmology observations. We demonstrate that COWaRD is realized, as an example, in a class of supersymmetric axion models and that future observations by the next generation Cosmic Microwave Background, Large Scale Structure, and 21-cm surveys can reveal the structure of the theory.
\end{abstract}

\date{\today}

\maketitle

\tableofcontents

\newpage

\section{Introduction}
\label{sec:intro}
In the last few decades, cosmology as a science has undergone a phase of accelerated expansion. With the advent of probes of the cosmic microwave background (CMB) as well as experiments mapping the distribution of matter in the Universe \cite{Beutler:2011hx,Heymans:2013fya,Ross:2014qpa,Alam:2016hwk,Kohlinger:2017sxk,Joudaki:2017zdt,Akrami:2018vks}, we are currently living in an era of unprecedented precision in our understanding of the Universe, its composition, and its history.

From all these observations, a ``standard'' picture has emerged as the best framework to explain the data: the \LC~model. In \LC, the majority of the matter in the Universe consists of non-relativistic (``cold'') particles whose dominant interactions among themselves and with the rest of the Universe (the Standard Model of particle physics and the cosmological constant $\Lambda$) are through gravity (``dark''). This component has then been appropriately dubbed ``cold dark matter'' (CDM).

Although the predictions of the \LC~model are in outstanding agreement with observational data, cosmological experiments are still far from unambiguously singling out \LC~as the only option in town. From the theoretical point of view, the mystery of dark matter (DM) and its properties (origin, mass, spin, interactions, etc.) are perhaps the most common playground for particle physicists interested in cosmology. Many different and well-motivated models have been proposed to explain it, often requiring extra components of the Universe beyond those of vanilla \LC. Furthermore, recent experiments hint to possible cracks in the \LC~paradigm. Local measurements of the expansion of the Universe~\cite{Abbott:2017smn,Riess:2019cxk,Wong:2019kwg,Freedman:2019jwv,Yuan:2019npk,Huang:2019yhh} and of the large-scale structure (LSS)~\cite{Heymans:2013fya,Kohlinger:2017sxk,Joudaki:2017zdt,Ade:2013lmv,Bohringer:2014ooa,MacCrann:2014wfa,Ade:2015fva,Joudaki:2016mvz,Hildebrandt:2016iqg,Boehringer:2017wvr,Troxel:2017xyo,Troxel:2018qll} are in disagreement with the predicted values from the fit of \LC~to the early Universe data \cite{Aghanim:2018eyx,Bianchini:2019vxp,Schoneberg:2019wmt}.

Perhaps the most well-known modifications to the standard \LC~model are warm dark matter (WDM) and dark radiation (DR), ubiquitous in well-motivated theoretical frameworks. The sterile neutrino is a popular WDM candidate and can be produced from the thermal scattering processes~\cite{Dodelson:1993je} or the non-thermal resonant conversion~\cite{Shi:1998km}. Axions can constitute DR~\cite{Ema:2017krp} or WDM~\cite{Co:2017mop, Harigaya:2019qnl} from parametric resonance, while moduli decays also produce relativistic axions~\cite{Acharya:2010zx, Cicoli:2012aq, Higaki:2012ar, Conlon:2013isa, Higaki:2013lra, Marsh:2015xka}. In supersymmetric theories, gravitinos can also constitute WDM and may acquire the relic abundance from scatterings during inflationary reheating~\cite{Moroi:1993mb, Rychkov:2007uq} or decays of the lightest observable superpartner (LOSP) after freeze-out~\cite{Feng:2003xh, Feng:2003uy, Feng:2004zu, Feng:2004mt, Feng:2008zza, Feng:2012rn}. Similarly, the axinos from reheating~\cite{Strumia:2010aa} or the LOSP decay~\cite{Rajagopal:1990yx, Covi:1999ty, Covi:2001nw, Brandenburg:2005he} can be WDM. In Twin Higgs models~\cite{Chacko:2005pe}, the twin photons and twin neutrinos also contribute to DR~\cite{Barbieri:2005ri, Barbieri:2016zxn,Chacko:2016hvu, Craig:2016lyx, Csaki:2017spo,Barbieri:2017opf,Harigaya:2019shz}. These references consider the presence of either WDM or DR.  Remarkably, precision cosmology can reveal information about these new states via gravitational effects alone although their direct interactions may otherwise be too weak for terrestrial experiments.

Certain well-motivated theories predict the simultaneous existence of WDM and DR.
In this paper, we point out that  precision cosmology can be even more powerful as to reveal the underlying theoretical structure from the correlated signals.
We consider a theory where late-decaying particles decay into nearly massless particles and stable massive particles. This predicts that stable particles are WDM that becomes non-relativistic just before the CMB
\begin{equation}
\label{eq:Tnr_fwdm_Neff}
T_{\rm nr} \simeq 5.5~{\rm eV} \left( \frac{f_{\rm wdm}}{\Delta N_{\rm eff}} \right),
\end{equation}
if both the effective number of relativistic species $\Delta N_{\rm eff}$ due to the nearly-massless particles and the fractional amount of such warm dark matter $f_{\rm wdm}$ are of the same order.
We call such models Common Origin of Warm and Relativistic Decay Products (COWaRD). If the decay products have sufficient interactions with the Standard Model, the required formation of light elements strongly constrains the decay to occur before Big-Bang Nucleosynthesis (BBN). In this case, the thermal interactions tend to equilibrate the decay products and remove the correlation in the signals. On the other hand, if the decay products are invisible, the decay can occur any time before recombination and both WDM and DR survive and leave imprints on the CMB via gravitational effects. These final states can be produced from two different decay channels or in a single decay process; the branching ratio determines the relative abundance for the former case, whereas Eq.~(\ref{eq:Tnr_fwdm_Neff}) is applicable to the latter. An example for the former case is the decay of the saxion into a pair of axions or a pair of axinos. For the latter case, examples include the axino/gravitino decay into the gravitino/axino and an axion $\tilde{a}/\tilde{G} \rightarrow \tilde{G}/\tilde{a} \ a$ as well as the saxion decay into the axino and the gravitino $s \rightarrow \tilde{a} \tilde{G}$. These examples can be generalized to a decay involving the gravitino and a Nambu-Goldstone boson and the fermionic superpartner. A different class of examples is the decay of a fermion into another and a Nambu-Goldstone boson, such as the majoron associated with lepton number breaking~\cite{Chikashige:1980ui} (see Ref.~\cite{10.1093/mnras/211.2.277} for the effect on structure). Due to the common origin, their relative abundance is fixed. For concreteness, we will interpret results in the scenario where $\tilde{a}/\tilde{G} \rightarrow \tilde{G}/\tilde{a} \ a$ is the origin of warm gravitino/axino dark matter and axion dark radiation. Here we comment on previous works on the decaying axino or gravitino. Ref.~\cite{Hasenkamp:2011em} focuses on DR aspects, while Refs.~\cite{Chun:1993vz,Ichikawa:2007jv} point out effects on matter spectrum, but no numerical analysis is performed. Contrary to a decay before recombination, Refs.~\cite{Hamaguchi:2017ihw, Bae:2019vyh} consider lifetimes as long as the age of the universe.

In this work, we investigate the impact this common origin of WDM and DR has on cosmology. We implement this model into the {\tt CLASS} Boltzmann solver~\cite{Blas:2011rf} and perform fits to observational data via Monte-Carlo Markov Chain scans of the available parameter space using {\tt MontePython}~\cite{Audren:2012wb,Brinckmann:2018cvx}. We find that this model marginally improves the fit to the data compared to \LC, and it establishes an anti-correlation between the amount of DR and the amplitude of the matter fluctuations at large scales, parameterized by $\sigma_8$. As a consequence of this, the detection of extra relativistic degrees of freedom by Stage-4 CMB surveys immediately imply, in the context of the COWaRD model, a value of $\sigma_8$ smaller than the prediction from \LC.

The paper is organized as follows. In Sec.~\ref{sec:review}, we review the qualitative explanations of how WDM and DR affect the CMB and the matter power spectrum. In Sec.~\ref{sec:WDM_DR}, we discuss explicit models that give rise to both WDM and DR via perturbative decays and derive the correlation between observables as a result of this common origin. In Sec.~\ref{sec:cosmo}, we fit the proposed models to the cosmological data sets and provide the interpretation of the results. Finally, we conclude in Sec.~\ref{sec:concl}.

\section{Cosmological Effects of Dark Radiation and Warm Dark Matter}
\label{sec:review}

In this section, we recall how dark radiation and warm dark matter affect cosmological observables. For reviews of this topic, see \eg \cite{Lesgourgues:2018ncw}.

\subsection{Dark Radiation}
Dark radiation is the radiation component of the universe in addition to the prediction of the Standard Model--photons and neutrinos. The abundance of the radiation in addition to photons is commonly parametrized by the effective number of neutrinos $N_{\rm eff}$,
\begin{align}
\rho_{\rm rad} - \rho_\gamma = N_{\rm eff} \frac{\pi^2}{30} \frac{7}{4 } \left(\frac{4}{11}\right)^{4/3} T_\gamma^4,
\end{align}
where $T_\gamma$ is the temperature of photons.
The Standard Model prediction is $N_{\rm eff} \simeq 3.046$~\cite{Mangano:2001iu,deSalas:2016ztq} and the DR abundance is parametrized by the deviation from it,
\begin{align}\label{eq:deltaN}
\Delta N_{\rm eff} = N_{\rm eff} - 3.046.
\end{align}

\begin{figure}
\centering
\includegraphics[width=0.7\textwidth]{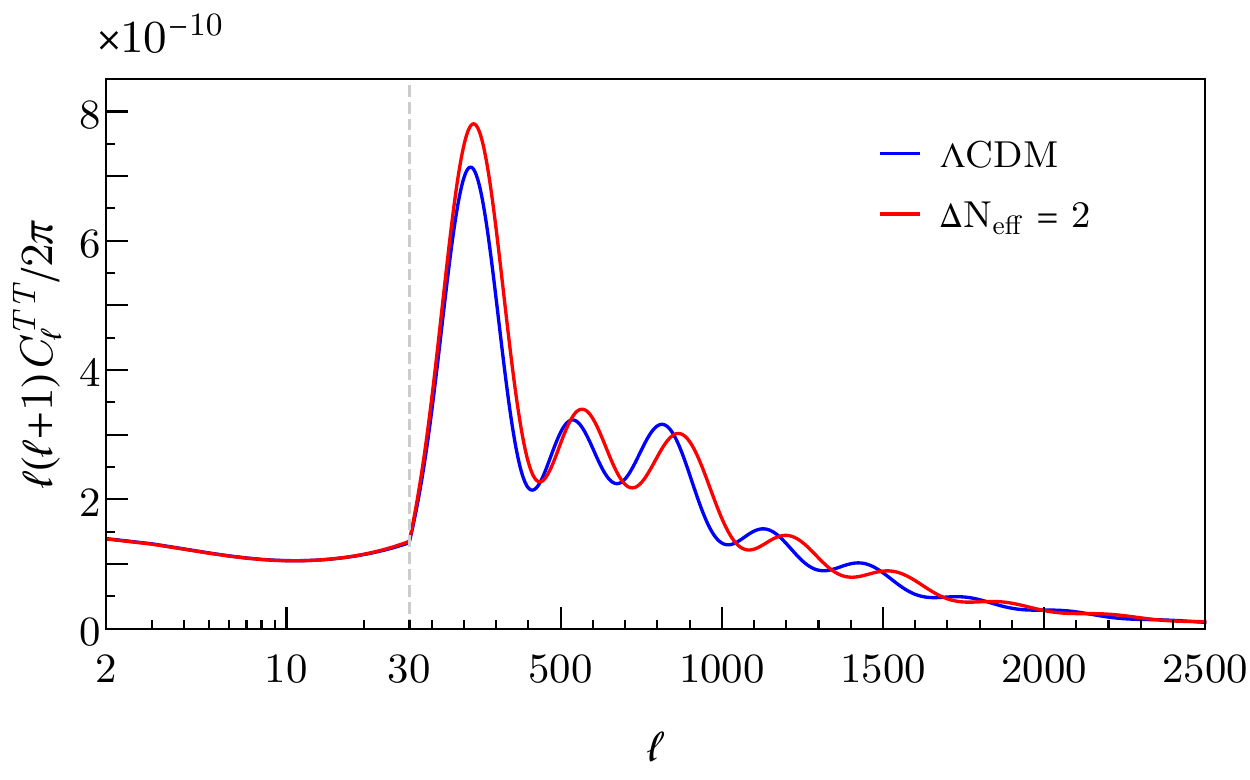}
\caption{The effect of dark radiation on the CMB $TT$ spectrum. The blue curve is for \LC~with best fit parameters from Planck 2018. In the red curve, two additional relativistic degrees of freedom are introduced to \LC. The scale of the horizontal axis changes from logarithmic to linear at $\ell = 30$.}
\label{fig:Neff}
\end{figure}

In Fig.~\ref{fig:Neff}, we show the CMB TT spectra for $\Lambda$CDM and $+\Delta N_{\rm eff}$ computed by {\tt CLASS}~\cite{Blas:2011rf}. Here we use the best fit cosmological parameters of $\Lambda$CDM determined by Planck 2018 TT+TE+EE+lensing and BAO data and fix $h = 0.6766$. The positive $\Delta N_{\rm eff}$ affects the CMB spectrum and the determination of the cosmological parameters in well understood ways \cite{Hou:2011ec,Chacko:2015noa,Follin:2015hya,Baumann:2015rya}, which we now briefly summarize.

Dark radiation increases the energy density of the universe for a given photon temperature and hence increases the expansion rate of the universe. The sound horizon, $r_s\sim c_s t$, around recombination becomes shorter, where $c_s$ is the sound speed. The angular position of the acoustic peaks of the CMB spectrum moves toward higher multipoles.
The angular position of the acoustic peaks is precisely measured by the observations of the CMB. To fix it at the correct position, the distance between us and the last scattering surface is required to be shorter. The distance is basically determined during the cosmological constant dominated era, and inversely proportional to $H_0$. Therefore, $H_0$ estimated from CMB observations becomes larger.

After fixing the positions of the peaks by increasing $H_0$, Silk damping~\cite{Silk:1967kq,Hu:1994jd,Hu:1996vq,Weinberg:2008zzc,Lesgourgues:2013qba} begins from lower multipoles (a larger expansion rate shortens the diffusion length, but this effect is subdominant). This can be partially compensated by 1) a larger spectral index $n_s$, which increases the primordial perturbation at small scales or 2) more baryons which shorten the diffusion length, although too large of a $\Delta N_{\rm eff}$ twists the overall shape. Furthermore, the polarization spectra have more power at high multipoles, which helps constrain $\DN$. Planck 2018 TT+TE+EE+lensing and BAO data set the constraint~\cite{Aghanim:2018eyx}
\begin{equation}
\label{eq:Neff_Planck2018}
\Delta N_{\rm eff} < 0.28~(95\% {\rm C.L.}) \ ,
\end{equation}
which Stage-4 ground-based CMB experiments are projected to push down to values of $\DN \approx 0.03$ \cite{Abazajian:2016yjj}.

The constraint (\ref{eq:Neff_Planck2018}) assumes that DR exists before the beginning of Big-Bang Nucleosynthesis (BBN).
In this case, since the expansion rate of the universe becomes larger around BBN, the neutron-proton conversion becomes relatively inefficient, making the neutron-to-proton ratio and hence the Helium fraction larger. Helium atoms have large binding energy than protons and recombine with electrons more efficiently. The scattering between CMB photons and electrons becomes less efficient, and the diffusion length of the CMB photons becomes longer. Silk damping begins from yet lower multipoles, strengthening the constraint on $\Delta N_{\rm eff}$. Thus, if DR is produced after BBN, the constraint on $\Delta N_{\rm eff}$ is weaker than the one in Eq.~(\ref{eq:Neff_Planck2018}).

If matter-radiation equality is delayed by a positive  $\Delta N_{\rm eff}$, the decay of the gravitational potential, which induces the acoustic oscillation, lasts longer. As a result, the amplitude of the first few acoustic peaks becomes larger.\footnote{The amplitude of the acoustic peaks at high multipoles is suppressed because of the shift of the peaks to higher multipoles and more effective Silk damping.} As an indirect consequence of this, dark radiation ends up affecting the matter spectrum too. In order to keep the time of matter-radiation equality fixed (which is well measured by CMB surveys), an increase in the amount of radiation must be compensated by a corresponding increase in the amount of matter. This in turn translates into a larger matter power spectrum.

\subsection{Warm Dark Matter}
Warm dark matter is a component of the universe whose energy density is dominated by a mass density around matter-radiation equality but has a large enough velocity dispersion to affect observations. Warm dark matter behaves in the same manner as cold dark matter at the background level, but differently at the perturbation level.

Warm dark matter freely streams and escapes from overdense regions, suppressing the perturbations at length scales shorter than the free-streaming length,
\begin{align}
\label{eq:FS}
%\lambda_{\rm FS} = \int \frac{v(t)}{a(t)}{\rm d}t \simeq \frac{v(t_{\rm eq})}{a(t_{\rm eq})} t_{\rm eq} \simeq 1~{\rm Mpc} \left( \frac{100~{\rm eV}}{T_{\rm nr}} \right),
\lambda_{\rm FS} = \int \frac{v(t)}{a(t)}{\rm d}t \simeq 4~{\rm Mpc} \left( \frac{100~{\rm eV}}{T_{\rm nr}} \right) \left( \frac{{\rm ln} (T_{\rm nr}/T_{\rm eq})}{5} \right),
\end{align}
where $T_{\rm nr}$ is the temperature around which WDM becomes non-relativistic.

The observed shortest scale in the CMB is about 10 Mpc. The observation of the CMB spectrum can constrain the abundance of WDM if $T_{\rm nr} \lesssim 10$ eV. For larger $T_{\rm nr}$, on the other hand, the effect of WDM can be observed in the matter power spectrum at small scales, which is measured by galaxy counting, Lyman-$\alpha$ power spectrum, and the Sunyaev-Zel'dovich and lensing effects on the CMB.

\subsection{$H_0$ and $\sigma_8$ tensions}\label{sec:tensions}

While undoubtedly very successful, the \LC~model is far from being singled out as the one and only option to explain the available cosmological observations. What is more, there are some indications that \LC~is not the whole story. In particular, the values of the Hubble expansion rate $H_0$ and the amplitude $\sigma_8$ of matter fluctuations at the scale of $8~\mpc$, as quantities derived from the fit of \LC~to the CMB data from the Planck satellite, are in tension with direct measurements of the same.

Indeed, direct measurements of the expansion rate of the Universe tend to yield a large value of $H_0$, whereas ``indirect'' measurements (\ie deductions from \LC~fits to data) favor smaller values \cite{Aghanim:2018eyx,Abbott:2017smn,Schoneberg:2019wmt,Freedman:2019jwv,Riess:2019cxk,Huang:2019yhh,Yuan:2019npk,Wong:2019kwg,Bianchini:2019vxp}. For some of these observations, the tension reaches $4-6 \sigma$ \cite{Bernal:2016gxb,Aylor:2018drw,Knox:2019rjx,Verde:2019ivm}. If this discrepancy is a sign of new physics and not of unaccounted for systematics, the most accepted models that can alleviate this tension are those that change the size of the comoving sound horizon at the time of recombination \cite{Bernal:2016gxb,Aylor:2018drw,Knox:2019rjx}. The most straightforward way to do this is by adding some extra energy density at early times~\cite{Hojjati:2013oya,Poulin:2018dzj,Poulin:2018cxd,Agrawal:2019lmo,Park:2019ibn,Kreisch:2019yzn,Smith:2019ihp}\footnote{However, see also \cite{Bringmann:2018jpr,Vattis:2019efj}.}, such as DR~\cite{Eisenstein:2004an,Buen-Abad:2015ova,Lesgourgues:2015wza,Chacko:2016kgg,Buen-Abad:2017gxg}. As is explained above, the data is highly sensitive to the behavior of the cosmological perturbations in this extra component, and the simple addition of DR cannot fully explain the tension, as can be seen from the result of Planck 2018~\cite{Aghanim:2018eyx}.

On the other hand, direct measurements of LSS observe a smaller value of $\sigma_8$ than the \LC~prediction from its fit to the CMB data \cite{Aghanim:2018eyx,Heymans:2013fya,Hildebrandt:2016iqg,Joudaki:2016mvz,Troxel:2018qll,Troxel:2017xyo,Ade:2015fva,Ade:2013lmv,Kohlinger:2017sxk,Joudaki:2017zdt,MacCrann:2014wfa,Bohringer:2014ooa,Boehringer:2017wvr}. The suppression of structure by WDM can explain the discrepancy. For works inspired by the $\sigma_8$ problem, see~\cite{Buen-Abad:2015ova,Lesgourgues:2015wza,Chacko:2016kgg,Canac:2016smv,Joudaki:2016kym,Ko:2016uft,Ko:2016fcd,Ko:2017uyb,Buen-Abad:2017gxg,Raveri:2017jto,Pan:2018zha,Chacko:2018vss,Buen-Abad:2018mas,Abazajian:2019ejt}.

\section{Warm Dark Matter and Dark Radiation as Decay Products}
\label{sec:WDM_DR}

\begin{figure}
\begin{center}
\includegraphics[width=0.96\linewidth]{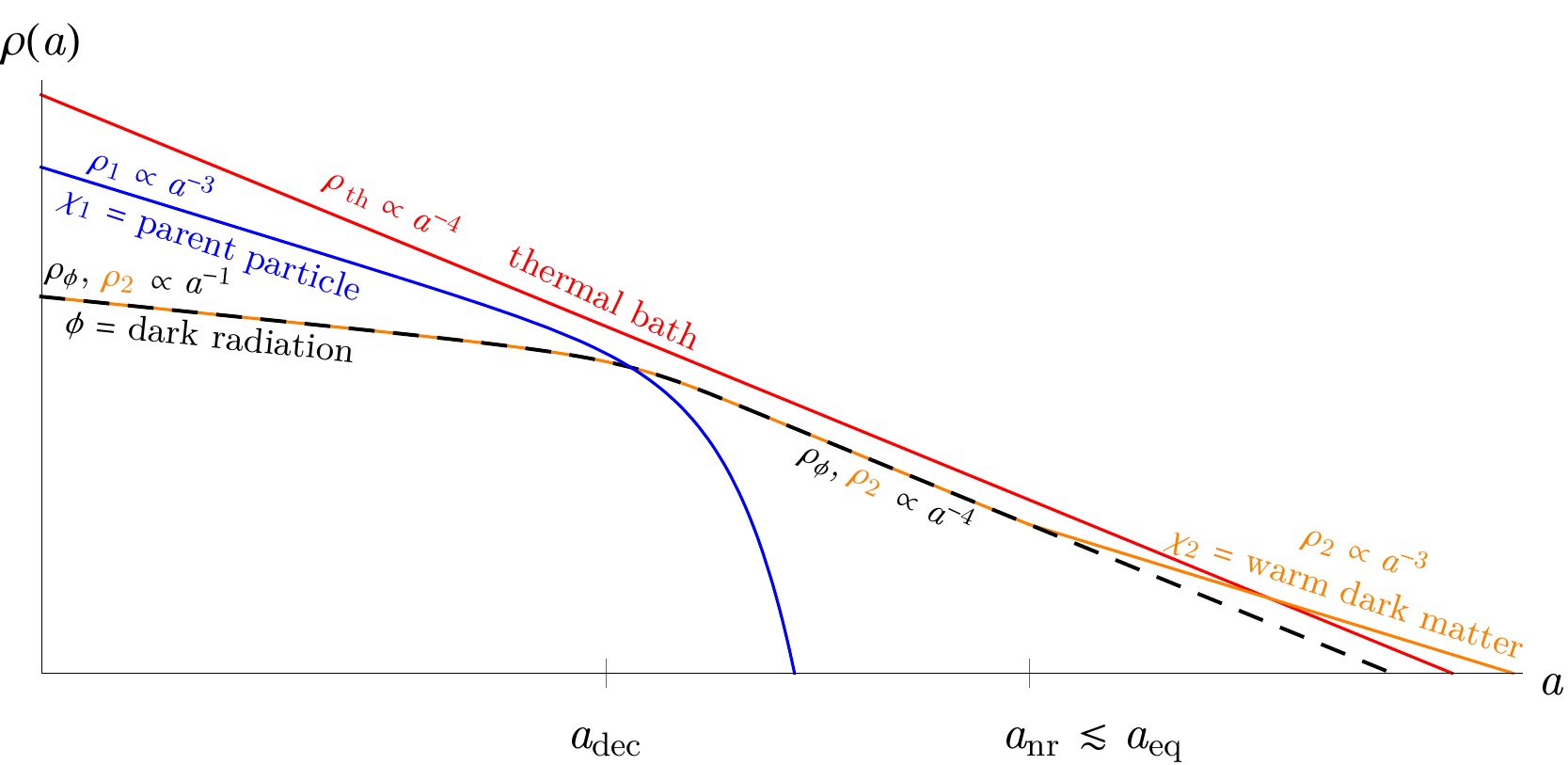}
\end{center}
\caption{The evolution of the energy densities $\rho$ with the scale factor $a$ on log-log scales for models of a common origin of WDM and DR. The red line shows the Standard Model thermal bath, while the blue curve is for the parent particle $\chi_1$ decaying into WDM $\chi_2$ (orange) and DR $\phi$ (black dashed curve). The decay occurs at $a_{\rm dec}$, while $a_{\rm nr}$--the scale factor when $\chi_2$ becomes non-relativistic--is close to matter-radiation equality $a_{\rm eq}$.}
\label{fig:rho}
\end{figure}

Here we consider the scenario where a late-decaying particle $\chi_1$ is produced in the early universe, and later decays into a stable particle $\chi_2$ and a stable and nearly massless particle $\phi$.
$\chi_2$ may behave as a WDM component or DR depending of the parameter space, and $\phi$ behaves as DR. In Fig.~\ref{fig:rho}, we demonstrate a cosmological evolution that leads to both WDM and DR. The existence of a sizable fraction of WDM and DR requires that 1) the decay occur when $\rho_1$ is $\mathcal{O}(0.1)$ fraction of the thermal bath energy density and 2) $\chi_2$ become non-relativistic shortly before matter-radiation equality. As we will show in this section, this scenario is realized in well-motivated models. On the other hand, if the requirement 1) is violated, the abundances of $\chi_2$ and $\phi$ are negligible or overproduced, while $\chi_2$ is cold dark matter or dark radiation if 2) is violated.

\subsection{Warmness and Abundances}
The Boltzmann equations of these particles read
\begin{subequations}
\begin{align}
\label{eq:rho_BoltEq}
\dot\rho_1 + 3 H \rho_1 & = - \Gamma \rho_1 \\
\dot\rho_2 + 4 H \rho_2 & =  \frac{1}{2} \Gamma \rho_1 \\
\dot\rho_\phi + 4 H \rho_\phi & =  \frac{1}{2} \Gamma \rho_1 \ .
\end{align}
\end{subequations}
For a radiation-dominated universe, the Hubble expansion rate is $H = 1/2t$ and the solutions are given by
\begin{align}
\label{eq:rho_BoltSol}
\rho_1 (t) & = \rho_1 (t_i) \left( \frac{t_i}{t} \right)^{3/2} e^{-\Gamma (t-t_i)} \\
\rho_{2, \phi} (t) & = \rho_1 (t_i) \left( \frac{t_i}{t} \right)^{3/2} \frac{\sqrt{\pi}}{4 \sqrt{t \Gamma}} \hspace{0.2 in} {\rm for} \hspace{0.2 in} t_i \ll \Gamma^{-1} \hspace{0.2 in} {\rm and} \hspace{0.2 in} t \gg \Gamma^{-1} ,
\end{align}
where $\rho_2 (t) = \rho_\phi (t)$ until $\chi_2$ becomes non-relativistic.
We parametrize the energy density of DR $\phi$ by the usual effective degrees of freedom $\Delta N_{\rm eff}$ so that
\begin{equation}
\rho_\phi =  \Delta N_{\rm eff} \frac{\pi^2}{30} \frac{7}{4} \left(\frac{4}{11}\right)^{{ \scalebox{1.01}{$\frac{4}{3}$} }} T^4 .
\end{equation}

The energy density of $\phi$ originates from the decay of $\chi_1$ at temperature $T_{\rm dec}$ and thus
\begin{equation}
\label{eq:DNeff_phi}
\Delta N_{\rm eff} = \frac{30}{\pi^2} \frac{4}{7} \left(\frac{11}{4}\right)^{{ \scalebox{1.01}{$\frac{4}{3}$} }} \frac{\rho_\phi}{T^4}
= \left(\frac{11}{4}\right)^{{ \scalebox{1.01}{$\frac{4}{3}$} }} \frac{ 4 \sqrt{2\pi}  g_*(T_{\rm dec})}{21} \frac{\rho_1/s}{T_{\rm dec}}   ,
\end{equation}
where $g_*$ is the effective number of degrees of freedom in the thermal bath, $s$ is the entropy density, and the decay temperature is defined by $H(T_{\rm dec}) = 1/2 t_{\rm dec} \equiv \Gamma$. We use $\rho_\phi(T_{\rm dec}) / s(T_{\rm dec}) = \rho_1(T_i)/s(T_i)  \sqrt{\pi/8}$ for an initial temperature $T_i \gg T_{\rm dec}$, which follows from the analytic solutions of the full Boltzmann equations given in Eq.~(\ref{eq:rho_BoltSol}). We further simplify the notation $\rho_1(T_i)/s(T_i)  = \rho_1/s$ because this ratio is a constant for any $T_i \gg T_{\rm dec}$.
Similarly, the number density of $\chi_2$ is equal to that of $\chi_1$ at the time of decay,
\begin{equation}
\label{eq:rho2s}
\left. \frac{\rho_2}{s} \right|_{T=0} =  \frac{m_2n_2}{s} = \frac{\rho_1}{s} \frac{m_2}{m_1} ,
\end{equation}
while $m_2n_2/s$ stays constant afterwards and stands for the energy density of $\chi_2$ at the zero temperature, \ie $\left. \rho_2 \right|_{T=0} $.
Lastly, the momentum of $\chi_2$ at $T_{\rm CMB} \simeq 0.25$ eV can be obtained from that at $T_{\rm dec}$
\begin{equation}
\label{eq:ptCMBm2T}
\left. \frac{p_2}{m_2} \right|_{\rm CMB} = \frac{p_{\rm dec}}{m_2} \frac{T_{\rm CMB}}{T_{\rm dec}} = \frac{m_1}{2 m_2} \frac{T_{\rm CMB}}{T_{\rm dec}} ,
\end{equation}
where $p_{\rm dec} =  m_1/2$ is a result of the four-momentum conservation assuming $m_2, m_\phi \ll m_1$. Equivalently, $\chi_2$ becomes non-relativistic at the temperature
\begin{equation}
\label{eq:TnrTdec}
T_{\rm 2,nr} = T_{\rm CMB} \left. \frac{m_2}{p_2} \right|_{\rm CMB} = \frac{2 m_2}{m_1} T_{\rm dec}.
\end{equation}

Now with Eqs.~(\ref{eq:DNeff_phi})-(\ref{eq:ptCMBm2T}), we can see the dependence of the velocity of $\chi_2$ on the amounts of dark radiation, $\Delta N_{\rm eff}$, and warm dark matter, $\left. \rho_2/s \right|_{T=0}$,
\begin{equation}
\label{eq:ptCMBm2}
\left. \frac{p_2}{m_2} \right|_{\rm CMB}  = \left(\frac{4}{11}\right)^{{ \scalebox{1.01}{$\frac{4}{3}$} }} \frac{21}{ 8 \sqrt{2\pi} g_*(T_{\rm dec})} \frac{ \Delta N_{\rm eff} T_{\rm CMB}}{\left. \rho_2 / s \right|_{T=0}} .
\end{equation}
Therefore, $\chi_2$ becomes non-relativistic at the temperature
\begin{align}
\label{eq:Tnr_fwdm_DNeff}
T_{2,{\rm nr}} & = \left(\frac{11}{4}\right)^{{ \scalebox{1.01}{$\frac{4}{3}$} }} \frac{ 8 \sqrt{2\pi} g_*(T_{\rm dec})}{21}  \frac{\left. \rho_2 / s \right|_{T=0}}{ \Delta N_{\rm eff}}  \simeq 5.5~{\rm eV} \left( \frac{f_{\rm wdm}}{\Delta N_{\rm eff}} \right) ,
\end{align}
which is interestingly just before the CMB epoch when $\Delta N_{\rm eff}$ and $f_{\rm wdm}$ are of the similar order. Here, $f_{\rm wdm} \equiv \rho_2 / \rho_{\rm DM}$ is the abundance of WDM $\chi_2$ in units of the total dark matter abundance today. However, this is merely a result of the fact that the CMB decoupling temperature is close to the temperature of matter-radiation equality. A common origin of WDM and DR necessarily implies that two species have the same number density and the same momentum when relativistic. It is when $\chi_2$ becomes non-relativistic that the two energy densities start to deviate. If this occurs at the CMB epoch, then the two species still have roughly the same energy densities at the CMB and furthermore $\Delta N_{\rm eff} \simeq \mathcal{O}(1)$ is equivalent to $\left. \rho_2 / s \right|_{T=0} \simeq \rho_{\rm DM}/s$ due to matter-radiation equality. COWaRD in principle involves three parameters, $T_{\rm dec}$, $f_{\rm wdm}$, and $\Delta N_{\rm eff}$. However, if the decay occurs well before the CMB, $T_{\rm dec}$ becomes irrelevant so $f_{\rm wdm}$ and $\Delta N_{\rm eff}$ are sufficient to fully describe COWaRD.

\subsection{Supersymmetric Axion Models}
\label{subsec:model}

In this section, we will study supersymmetric axion theories because both supersymmetry and the axion are well motivated to solve outstanding issues in the Standard Model. Supersymmetry provides a solution to the electroweak hierarchy problem~\cite{Maiani:1979cx,Veltman:1980mj,Witten:1981nf,Kaul:1981wp}, a framework for precise gauge coupling unification~\cite{Dimopoulos:1981yj,Dimopoulos:1981zb,Sakai:1981gr,Ibanez:1981yh,Einhorn:1981sx,Marciano:1981un}, and stable massive particles as dark matter candidates~\cite{Witten:1981nf,Pagels:1981ke,Goldberg:1983nd}. On the other hand, a vanishing neutron electric dipole moment calls for an explanation, known as the strong CP problem~\cite{tHooft:1976rip}. A dynamical solution is provided by the Peccei-Quinn mechanism~\cite{Peccei:1977hh,Peccei:1977ur}, where a hypothetical particle called the axion~\cite{Weinberg:1977ma,Wilczek:1977pj} relaxes to a CP-conserving minimum in the potential and cancels the bare strong CP angle in the theory. The axion is a dark matter candidate~\cite{Preskill:1982cy,Abbott:1982af,Dine:1982ah}, and furthermore may explain the baryon asymmetry of the Universe~\cite{Kuzmin:1992up,Servant:2014bla,Kusenko:2014uta,Ipek:2018lhm,Co:2019wyp}. In supersymmetric axion theories, the Peccei-Quinn symmetry breaking scale can be determined by the dimensional transmutation through the running of the soft mass of the Peccei-Quinn symmetry breaking field~\cite{Moxhay:1984am} or by balance between a negative soft mass term and a higher dimensional term of the Peccei-Quinn symmetry breaking field~\cite{Murayama:1992dj}.

In the context of supersymmetric axion theories, we explore explicit models that give rise to both warm dark matter $\chi_2$ and dark radiation $\phi$ via a perturbative decay $\chi_1 \rightarrow \chi_2 + \phi$. One interesting scenario is the decay of axinos to gravitinos or vice versa, and both decay directions come with axions. The WDM candidate $\chi_2$ is then the gravitino~\cite{Witten:1981nf} or the axino~\cite{Rajagopal:1990yx,Covi:1999ty}, while the axion produced from the decay behaves as dark radiation $\phi$. The decay can safely occur after BBN since the decay products have negligible interactions with the Standard Model. Previous works~\cite{Chun:1993vz,Ichikawa:2007jv} provide qualitative discussions, while we give a fully quantitative analysis in Sec.~\ref{sec:cosmo}. In addition, Refs.~\cite{Chun:1993vz,Ichikawa:2007jv} consider the spectrum where the axino is (much) lighter than the gravitino. The axino and gravitino masses are highly model dependent~\cite{Goto:1991gq, Chun:1992zk, Chun:1995hc, Kim:2012bb, Kawasaki:2013ae} and can also be of the same order in realistic models.
For example, the axino can acquire a mass at one loop level via the heavy (s)quarks in KSVZ models, which allows the axino to be heavier than the gravitino in gauge mediation when the PQ breaking scale is below the messenger scale. On the other hand the axino can be massless at the tree level for no-scale supersymmetry, resulting in the axino much lighter than the gravitino. Due to this model uncertainty, we treat the gravitino and axino masses as free parameters.

In what follows, we will show that the decay rate with well-motivated ranges of parameters leads to a relativistic species as well as a warm species that becomes non-relativistic shortly before the CMB epoch. The amounts of produced DR and WDM depend on the abundance of the parent particle at the time of the decay as shown in Eqs.~(\ref{eq:DNeff_phi}) and (\ref{eq:rho2s}). The abundance is determined by physics at the high temperatures, such as scatterings during inflationary reheating and freeze-out processes around the TeV scales. We defer the discussion of parent particle's relic abundance to App.~\ref{sec:prod_parent} and simply assume the desired abundance in this section.

We will first focus on the decay of axinos to gravitinos, which is relevant when the axino mass $m_{\tilde{a}}$ is larger than the gravitino mass $m_{3/2}$, and the decay rate is given by \cite{Hamaguchi:2017ihw}:
\begin{equation}
\Gamma_{\tilde{a} \rightarrow \tilde{G} a} = \frac{m_{\tilde{a}}^5}{96\pi m_{3/2}^2 M_{\rm Pl}^2} \left(1- \frac{m_{3/2}}{m_{\tilde{a}}} \right)^2\left(1- \left( \frac{m_{3/2}}{m_{\tilde{a}}} \right)^2\right)^3 ,
\end{equation}
where the reduced Planck mass $M_{\rm Pl} = 2.4 \times 10^{18}$ GeV. The dependence on the gravitino mass can be understood in the following way. The non-zero axino mass requires supersymmetry breaking, since it is a superpartner of the nearly massless axion. For a smaller gravitino mass, the supersymmetry breaking scale is smaller, and hence a larger coupling between the axino and the supersymmetry breaking field is required to realize the axino mass. The supersymmetry breaking field contains the spin-$1/2$ components of the gravitino, and the axino-axion-gravitino coupling is larger, enhancing the decay rate. The formula breaks down when $m_{3/2}M_{\rm Pl} < m_{\tilde{a}}^2$, which does not occur in the parameter space we consider. The decay of axinos dominantly occurs when the decay rate is comparable to the Hubble rate, occurring at a temperature
\begin{equation}
\label{eq:TdecAxino}
T_{\tilde{a} \rightarrow \tilde{G} a} =  \left( \frac{5}{2 g_*(T_{\tilde{a} \rightarrow \tilde{G} a})} \right)^{{ \scalebox{1.01}{$\frac{1}{4}$} }} \frac{m_{\tilde{a}}^{5/2}}{4\pi \sqrt{M_{\rm Pl}} m_{3/2}}
\simeq  10 \, {\rm eV}  \left( \frac{m_{\tilde{a}}}{9~\GeV} \right)^{{ \scalebox{1.01}{$\frac{5}{2}$} }}  \left( \frac{1~\GeV}{m_{3/2}} \right) \left( \frac{4}{ g_*(T_{\tilde{a} \rightarrow \tilde{G} a})} \right)^{{ \scalebox{1.01}{$\frac{1}{4}$} }} ,
\end{equation}
assuming a radiation-dominated epoch.
With the decay temperature from Eq.~(\ref{eq:TdecAxino}), one only needs to specify the abundance of axino at the time of decay, $\rho_{\tilde{a}} (T_{\rm dec})/s(T_{\rm dec})$, in order to obtain the amounts of the axion $\Delta N_{\rm eff}$ from Eq.~(\ref{eq:DNeff_phi}) and of the gravitino $\left. \rho_{3/2} / s \right|_{T=0}$ from Eq.~(\ref{eq:rho2s}). Using Eq.~(\ref{eq:ptCMBm2T}), we can now estimate when the gravitino becomes non-relativistic,
\begin{equation}
\label{eq:warm_gravitino}
T_{\tilde{G}, {\rm nr}} =  \frac{2 m_{3/2}}{m_{\tilde{a}}} T_{\tilde{a} \rightarrow \tilde{G} a}  \simeq 2~{\rm eV} \left( \frac{m_{\tilde{a}}}{9~\GeV} \right)^{{ \scalebox{1.01}{$\frac{3}{2}$} }} \left( \frac{4}{ g_*(T_{\tilde{a} \rightarrow \tilde{G} a})} \right)^{{ \scalebox{1.01}{$\frac{1}{4}$} }} .
\end{equation}
The result interestingly only depends on the axino mass but not the gravitino mass. This means that the gravitino can be arbitrarily light, while the  gravitino is equally warm during the CMB epoch for an axino mass of order 10-100 GeV. A smaller gravitino mass implies that the gravitino is hotter at the time of the axino decay, while the decay itself occurs earlier so that a longer era of redshift exactly compensates a larger initial gravitino momentum.

Now we discuss the alternative mass spectrum where the gravitino is heavier than the axino. In this case, the gravitino decay sources axino WDM and axion DR with a rate \cite{Hamaguchi:2017ihw}:
\begin{equation}
\Gamma_{\tilde{G} \rightarrow \tilde{a} a} = \frac{m_{3/2}^3}{192 \pi M_{\rm Pl}^2} \left(1- \frac{m_{\tilde{a}}}{m_{3/2}} \right)^2\left(1- \left( \frac{m_{\tilde{a}}}{m_{3/2}} \right)^2\right)^3 .
\end{equation}
In a radiation-dominated epoch, the gravitino decays at the temperature
\begin{equation}
\label{eq:TdecGravitino}
T_{\tilde{G} \rightarrow \tilde{a} a} = \left( \frac{5}{8 g_*(T_{\tilde{G} \rightarrow \tilde{a} a})} \right)^{{ \scalebox{1.01}{$\frac{1}{4}$} }} \frac{m_{3/2}^{3/2}}{4\pi \sqrt{M_{\rm Pl}}}
\simeq  10 \, {\rm eV} \left( \frac{m_{3/2}}{50~\GeV} \right)^{{ \scalebox{1.01}{$\frac{3}{2}$} }} \left( \frac{4}{ g_*(T_{\tilde{G} \rightarrow \tilde{a} a})} \right)^{{ \scalebox{1.01}{$\frac{1}{4}$} }} ,
\end{equation}
and the axino becomes non-relativistic at
\begin{equation}
\label{eq:warm_axino}
T_{\tilde{a}, {\rm nr}} =  \frac{2 m_{\tilde{a}}}{m_{3/2}} T_{\tilde{G} \rightarrow \tilde{a} a}  \simeq 2~{\rm eV} \left( \frac{m_{\tilde{a}}}{5~\GeV} \right) \left( \frac{m_{3/2}}{50~\GeV} \right)^{{ \scalebox{1.01}{$\frac{1}{2}$} }} \left( \frac{4}{ g_*(T_{\tilde{G} \rightarrow \tilde{a} a})} \right)^{{ \scalebox{1.01}{$\frac{1}{4}$} }} .
\end{equation}
In this mass spectrum, the decay occurs before the CMB as long as $m_{3/2} > \mathcal{O}(10)$ GeV, while the axino mass needs to be $m_{\tilde{a}} \simeq \mathcal{O}(10)~\GeV \left( 30~\GeV / m_{3/2} \right)^{1/2}$ to obtain WDM.

\subsection{Parameter Space with Suppression of $\sigma_8$}

\begin{figure}
\centering
\includegraphics[width=0.49\textwidth]{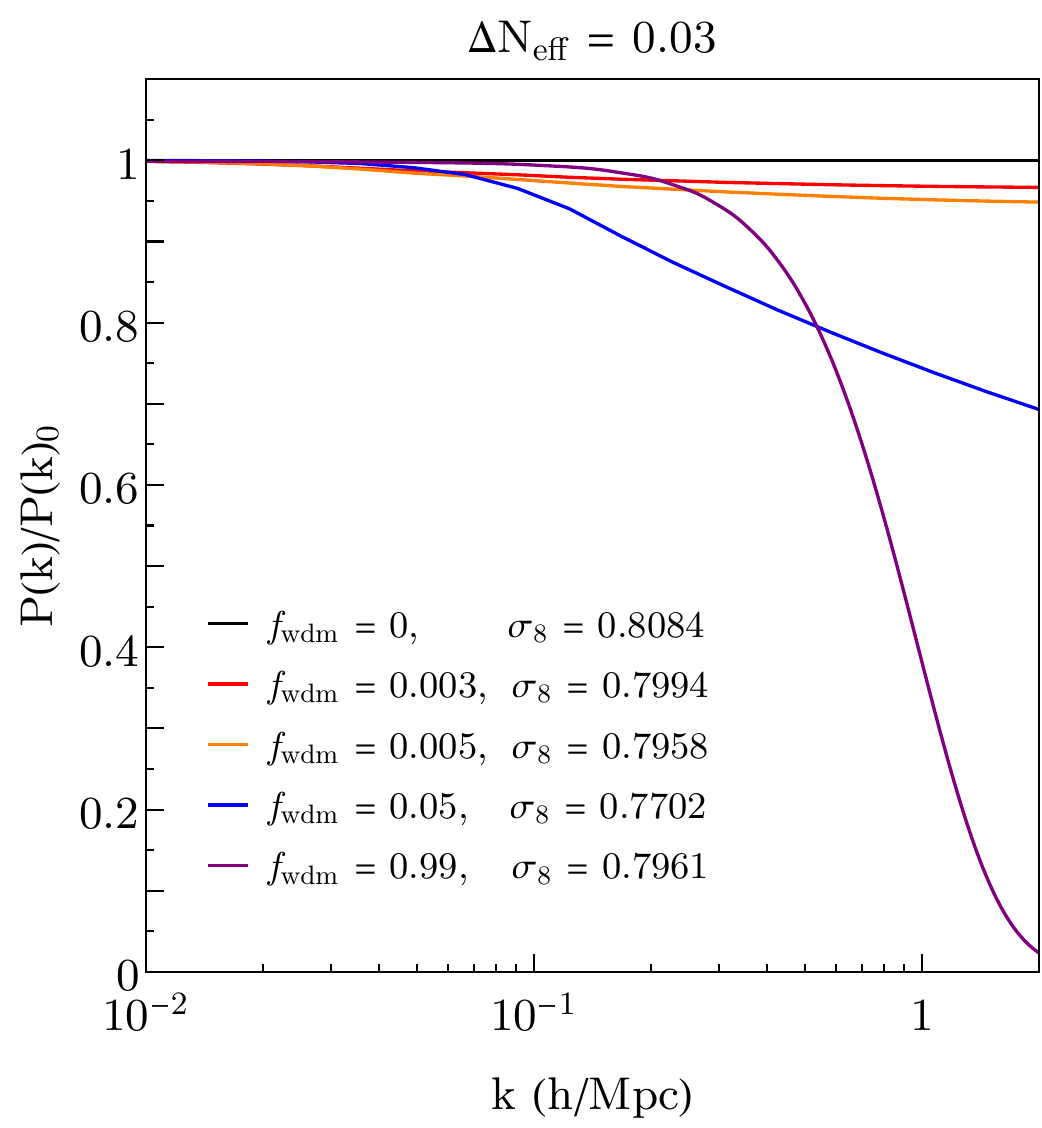}
\includegraphics[width=0.5\textwidth]{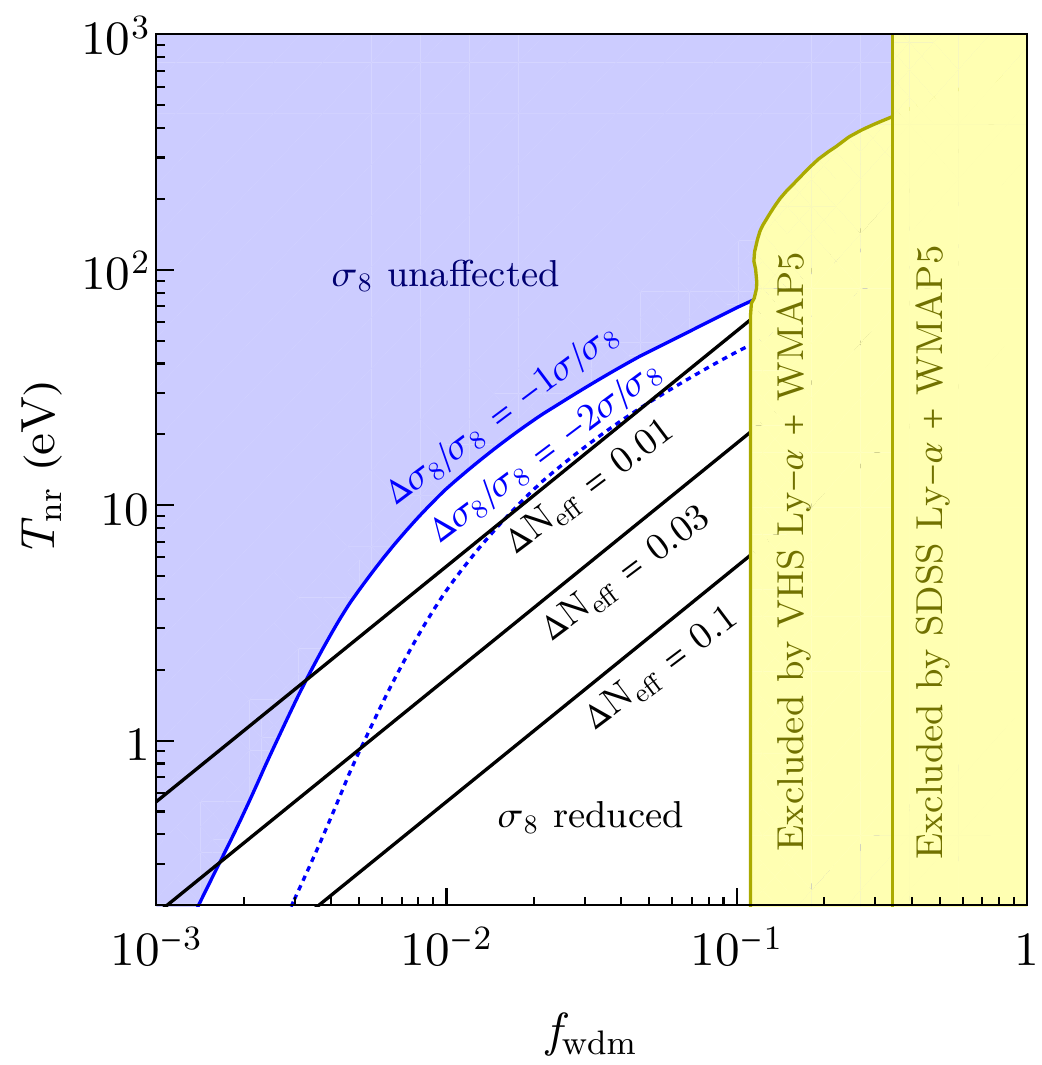}
\caption{The suppression of the matter spectrum due to both WDM and DR as decay products. {\bf Left panel}: the relative change compared to cold dark matter in the matter power spectra for various fractional amounts of WDM $f_{\rm wdm}$. {\bf Right panel}: below the solid (dotted) blue contours $\sigma_8$ is significantly suppressed, \ie more than $1 (2) \sigma/\sigma_8$ compared to \LC~extended by the same amount of $\Delta N_{\rm eff}$ as in the COWaRD model (black contours). The yellow region is excluded by Lyman-$\alpha$ constraints~\cite{Viel:2004bf, McDonald:2004eu, McDonald:2004xn, Boyarsky:2008xj}.}
\label{fig:Pk}
\end{figure}

The existence of WDM and DR leads to a suppressed matter power spectrum. The effect of WDM on cosmological observables depends on both its abundance and velocity around the CMB epoch.
For a given $\Delta N_{\rm eff}$, according to Eq.~(\ref{eq:Tnr_fwdm_DNeff}), the abundance and warmness of $\chi_2$ are inversely correlated. In the left panel of Fig.~\ref{fig:Pk}, we fix $\Delta N_{\rm eff}=0.03$ and compute the matter power spectrum as a function of the wavenumber for various values of $f_{\rm wdm}$.
We normalize the spectrum by that of $f_{\rm wdm}=0$. For a fixed $\Delta N_{\rm eff}$, a larger $f_{\rm wdm}$ implies colder WDM, whose correlation is given by Eq.~(\ref{eq:Tnr_fwdm_DNeff}). As $f_{\rm wdm}$ is increased, the matter spectrum is suppressed more. Since the free-streaming length becomes shorter, the suppression occurs only for smaller scales. For $f_{\rm wdm} \sim 1$, the impact on $\sigma_8$ is hence minor, but the power at $k \gtrsim 1$ h/Mpc is significantly suppressed.

We scan over the COWaRD parameters $f_{\rm wdm}$ and $T_{\rm nr}$ and examine the impact on $\sigma_8$. To be rigorous, one should also scan over other cosmological parameters to find the preferred range of parameters for a given $(f_{\rm wdm},T_{\rm nr})$, and predict $\sigma_8$. However, since WDM dominantly affects the prediction of $\sigma_8$ without affecting the CMB spectrum much, we can approximately infer the impact in the following way. We compute the ratio of $\sigma_8$ obtained in COWaRD containing both WDM and DR to that of \LC~extended with the same value of $\Delta N_{\rm eff}$ that COWaRD predicts, for fixed \LC~cosmological parameters. Using this methodology in the right panel of Fig.~\ref{fig:Pk}, we identify the region of parameter space below the solid (dotted) blue contours that leads to a fractional change $(\sigma_8^{\rm COWaRD} - \sigma_8) / \sigma_8$ more than $1(2) \sigma / \sigma_8 \simeq 0.75 (1.5) \%$ where $\sigma$ and $\sigma_8$ are the uncertainty and the measurement of $\sigma_8$ by Planck 2018. The yellow region is excluded by Lyman-$\alpha$ constraints~\cite{Boyarsky:2008xj}, which are based on the analyses presented by Viel-Haehnelt-Springel~\cite{Viel:2004bf} and for the SDSS experiment~\cite{McDonald:2004eu, McDonald:2004xn}.

\begin{figure}
\centering
\includegraphics[width=0.49\textwidth]{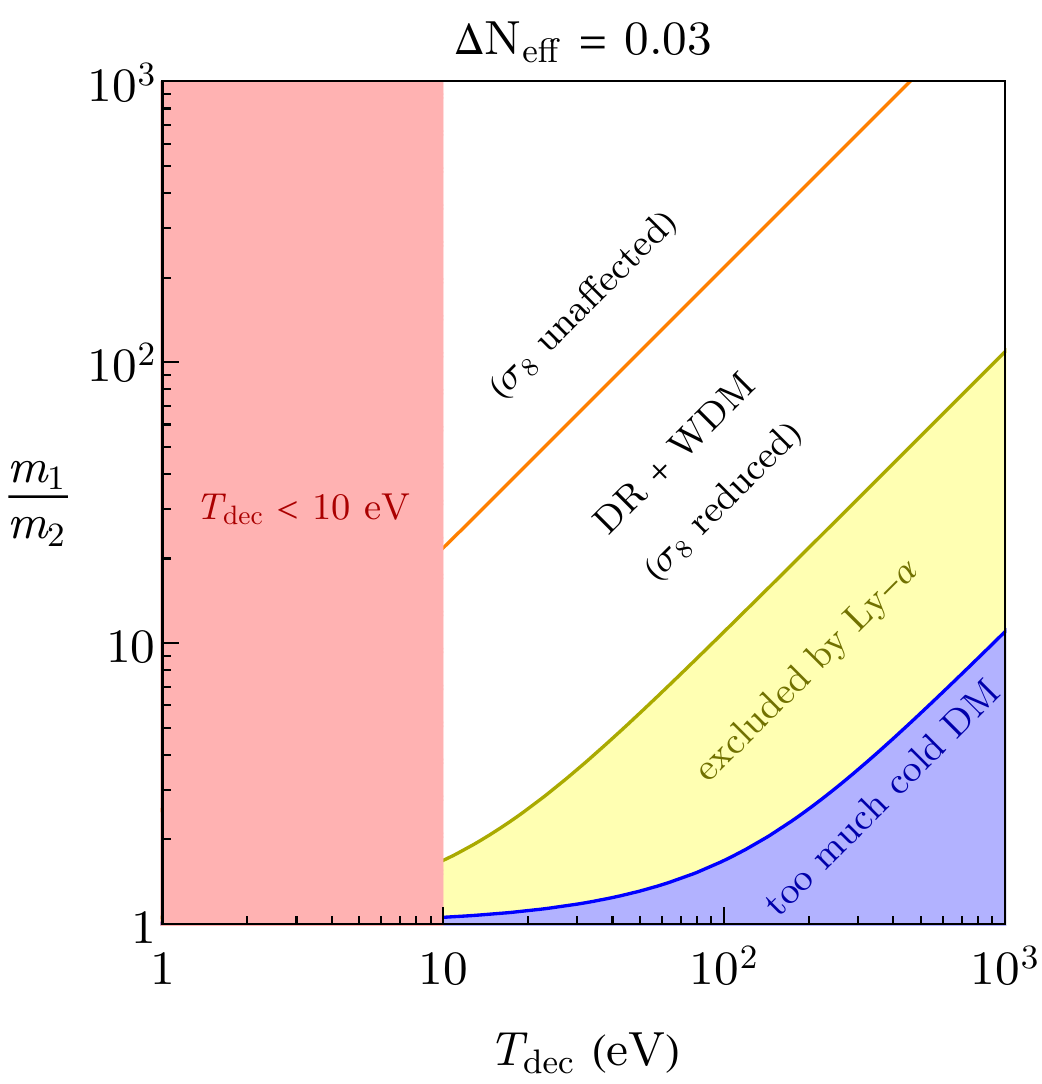}
\includegraphics[width=0.5\textwidth]{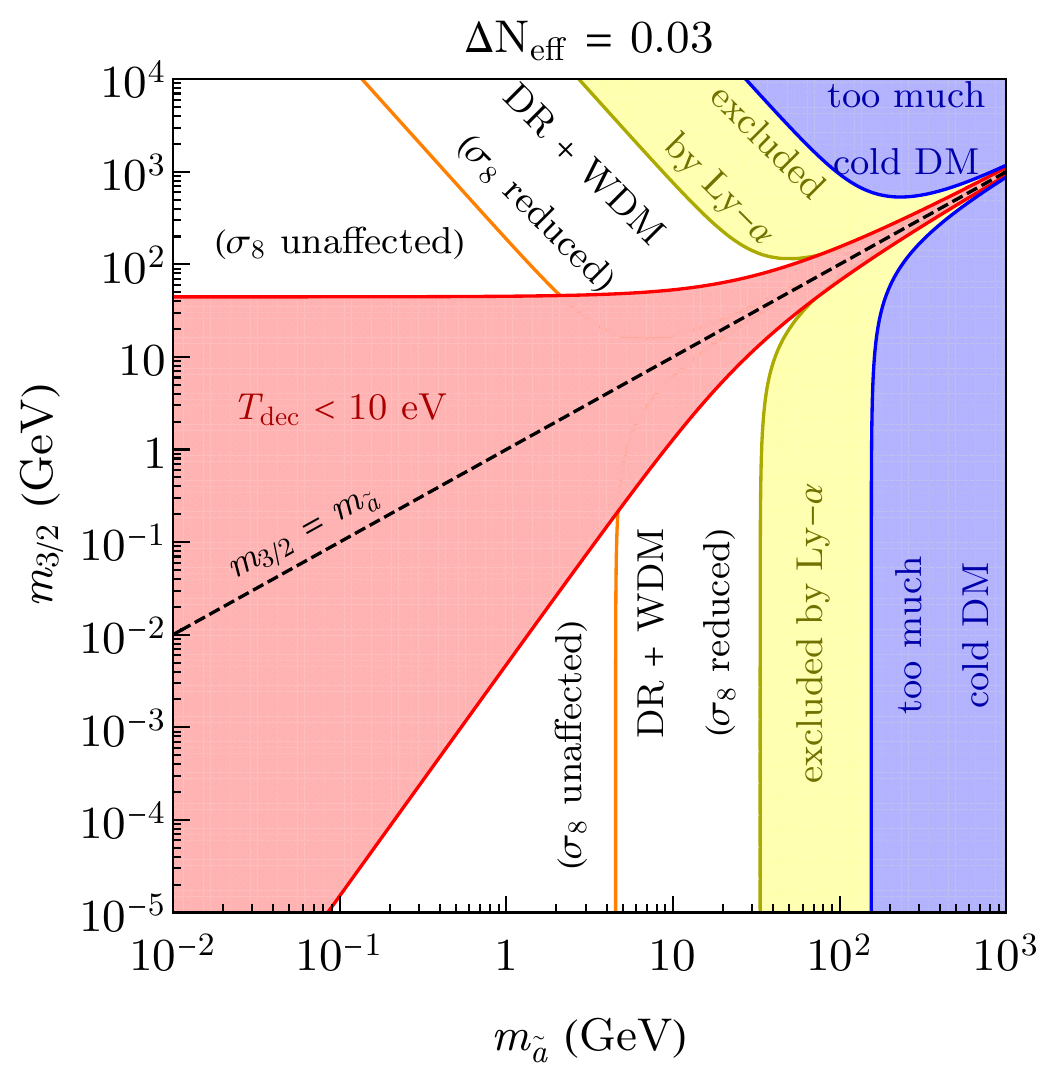}
\caption{The region between the orange and yellow curves leads to a significant suppression of the matter power spectrum at 8 Mpc scale $\sigma_8$, in the parameter space of {\bf left panel}: the mass ratio $m_1 / m_2$ between the parent and daughter particles and the decay temperature $T_{\rm dec}$, {\bf right panel}: the gravitino mass $m_{3/2}$ and the axino mass $m_{\tilde{a}}$.}
\label{fig:s8_effect}
\end{figure}

In Fig.~\ref{fig:s8_effect}, for a fixed $\Delta N_{\rm eff}=0.03$, we show the parameter space in particle physics models where $\sigma_8$ is significantly suppressed with the criterion defined in the methodology we used in Fig.~\ref{fig:Pk}. In the left panel, $m_1/m_2$ is the ratio of the parent and daughter particle masses and $T_{\rm dec}$ is the temperature at which the decay occurs. Warm dark matter and dark radiation lead to a significant ($2\sigma$) reduction in $\sigma_8$ in the region below the orange line, where $2\sigma$ is defined above. The region above the orange line on the other hand has $\sigma_8$ unaffected because the WDM abundance is too small despite WDM being very hot. The yellow region is excluded by the observations Lyman-$\alpha$ forests as also shown in Fig.~\ref{fig:Pk}.
In the red region, the decay occurs after the smallest observed scale in the CMB enters the horizon, and the CMB power spectrum is expected to be distorted in a way different from what a simple addition of WDM and DR does. Although the region may be viable, it is beyond the scope of our present analysis.
The blue region leads to too much cold dark matter as inferred by Eqs.~(\ref{eq:ptCMBm2T}) and (\ref{eq:ptCMBm2}) for a fixed $\Delta N_{\rm eff}$. The right panel is similar except that the parameter space is for the concrete supersymmetric axino/gravitino models elaborated in Sec.~\ref{subsec:model}. The black dashed line divides the region into one with a gravitino heavier than the axino (above) and the other with the opposite mass hierarchy. Interestingly, axinos lighter than $\mathcal{O}(30)$ GeV and gravitinos with any mass can lead to the prediction of COWaRD--a nonzero $\Delta N_{\rm eff}$ and a suppressed $\sigma_8$.

\section{Implications for Cosmology}
\label{sec:cosmo}

As stated previously, there is currently a plethora of cosmological observations that constrain any extensions to the Standard Model of cosmology \LC. In the model of COWaRD, the main two ingredients beyond those of vanilla \LC~are the WDM and DR components, which have a common origin and are related by \Eq{eq:ptCMBm2}. In this section, we describe the implementation of COWaRD into a code for the evolution of its cosmological perturbations, as well as its fit to the available cosmological data, and explain the results.

\subsection{Setting and Parameterization}

We first note that there is a wide array of values for the time of decay $t_\dec$, deep inside the radiation-dominated era, to which the available cosmological probes measuring the CMB and LSS are not sensitive. In terms of the corresponding scale factor $a_\dec$, this range is given by $a_{\rm BBN} \ll a_\dec \ll a_{\rm probed}$, where $a_{\rm BBN} \sim 10^{-8}$ is the scale factor at the time of BBN, and $a_{\rm probed} \sim 10^{-5}$ corresponds to the earliest time at which the smallest scales measured by the cosmological probes enter the horizon. Indeed for $a_\dec$ satisfying these inequalities there are ({\it i}) no effects of $\DN$ on BBN to speak of, and ({\it ii}) no consequences, coming from the decay process itself, on the cosmological observables. We corroborated that indeed the fits to cosmological data are insensitive to $a_\dec$ within this range and therefore, without loss of generality, we fix $a_\dec = 10^{-7}$. A more general implementation of this model, where the decay into WDM and DR is exactly modeled, at both the background and perturbations level and for any value of $a_\dec$ whatsoever, is beyond the scope of this paper and left to future work.

To calculate the impact that our COWaRD model has on cosmology we use the Boltzmann code {\tt CLASS} \cite{Blas:2011rf}. We take advantage of {\tt CLASS}'s options for the inclusion of dark radiation and non-cold (\ie warm) dark matter and modify them to suit our purposes. More concretely, since within COWaRD both DR and WDM have a common origin, we implement Eqs. (\ref{eq:TnrTdec}) and (\ref{eq:Tnr_fwdm_DNeff}), which relate the properties of both components. In addition, to solve for the evolution of the WDM perturbations, we implemented in {\tt CLASS} the WDM phase space distribution function $f(q)$, where $q=ap$ is the comoving momentum of the particles. From the Boltzmann equation with a decay term, we find $f(q)$ to be given by
\begin{equation}
  f(q) = \frac{\rho_{2,0}}{m_2} \frac{2\pi^2}{q q_\dec^2} ~ e^{-\frac{q^2}{2 q_\dec^2}} ~ \Theta(p_\dec - q) \ ,
\end{equation}
where we define $q_\dec \equiv a_\dec p_\dec$ and $\rho_{2,0} \equiv \rho_2 \vert_{T=0}$ is today's value of the energy density in the massive daughter particles.

As stated above, we fix $a_\dec = 10^{-7}$. This simplified version of the COWaRD model is then described by two parameters--the amounts of WDM and of DR. We denote these quantities respectively by $f_\wdm \equiv \frac{\rho_{2,0}}{\rho_{\rm DM,0}} = \frac{\rho_{2,0}}{\rho_{2,0} + \rho_{\cdm,0}}$, the fraction of DM today that contributed as WDM; and by $\DN$, given by \Eq{eq:deltaN}. The warmness of WDM, which is related to the time at which the WDM becomes non-relativistic, can be obtained from these two parameters with \Eq{eq:ptCMBm2}. Note that one can alternatively choose to parameterize the COWaRD model in terms of more fundamental particle physics parameters, such as the $m_2/m_1$ mass ratio, instead of either $\DN$ or $f_\wdm$. We decide to keep $\DN$ and $f_\wdm$ as our parameters since ({\it i}) their cosmological interpretation is more straightforward, ({\it ii}) they connect to established DR and WDM literature, as well as to published observational bounds, and ({\it iii}) allow for a more efficient exploration of the parameter space by MCMC methods near these bounds. Furthermore, we remind the reader that $m_2/m_1$ can be derived from $\DN$ and $f_\wdm$ using \Eqs{eq:TnrTdec}{eq:Tnr_fwdm_DNeff}.

To study the fit of the COWaRD model to current cosmological data we perform parameter scans over $\DN$ and $f_\wdm$, in addition to the six standard \LC~parameters. We make these scans using the Monte Carlo Markov Chain (MCMC) code {\tt MontePython} version 3.2 \cite{Audren:2012wb,Brinckmann:2018cvx} with the Metropolis-Hastings algorithm\footnote{We have included a joint prior on SZ nuisance parameters in the Planck 2018 likelihoods, according to Eq. (23) of \cite{Aghanim:2019ame}. This prior is now part of the newly released {\tt MontePython 3.3}.}. Our priors on $\DN$ and $f_\wdm$ are linear and only require these parameters to be non-negative. Finally, we include one massive neutrino of mass $m_\nu = 0.06~\EV$.

\subsection{The Data}

For our MCMC analysis, we use the following data sets:
\begin{itemize}
    \item {\bf CMB}: We use the published Planck 2018 TT+TE+EE and lensing data \cite{Akrami:2018vks}.
    \item {\bf BAO}: We include measurements of $D_V/r_\mathrm{drag}$ by 6dFGS at $z = 0.106$ \cite{Beutler:2011hx}, by SDSS from the MGS galaxy sample at $z = 0.15$ \cite{Ross:2014qpa}, and by BOSS from the CMASS and LOWZ galaxy samples of SDSS-III DR12 at $z = 0.2 - 0.75$ \cite{Alam:2016hwk}.
    \item {\bf Pantheon}: We also fit to the Pantheon set of type Ia supernovae (SN Ia) \cite{Scolnic:2017caz}, which consists of 1048 luminosity distances in the redshift range $0.01 < z < 2.3$. We include the nuisance parameter $M$ for the absolute magnitude of the SN Ia.
\end{itemize}

Combining the Planck satellite's measurements of the CMB with BAO and Pantheon has the advantage of narrowing down the values of different parameters, including the sound horizon and the evolution of the energy density of the Universe at late times. There are of course other data sets, perhaps the most prominent being the local measurements of the expansion of the Universe and of the large scale structure \cite{Aghanim:2018eyx,Abbott:2017smn,Schoneberg:2019wmt,Freedman:2019jwv,Riess:2019cxk,Huang:2019yhh,Yuan:2019npk,Wong:2019kwg,Bianchini:2019vxp,Heymans:2013fya,Hildebrandt:2016iqg,Joudaki:2016mvz,Troxel:2018qll,Troxel:2017xyo,Ade:2015fva,Ade:2013lmv,Kohlinger:2017sxk,Joudaki:2017zdt,MacCrann:2014wfa,Bohringer:2014ooa,Boehringer:2017wvr}. However, as we discussed in \Sec{sec:tensions}, some of these measurements seem to be in tension with measurements from the Planck satellite, and thus we do not include them in our MCMC analysis.

\subsection{Results}
We do our numerical analysis for both the standard \LC~model and the COWaRD model, fitting their parameters to the cosmological data described in the previous section. We find that the COWaRD model only very marginally improves the fit to the data, with a $\Delta \chi^2_{\rm eff} \sim {\rm few}$ (see \Tab{tab:allchi2} in \App{appB}). However, this fit sheds light on some important cosmological predictions of the COWaRD model.

\begin{figure}
\centering
\includegraphics[width=0.75\textwidth]{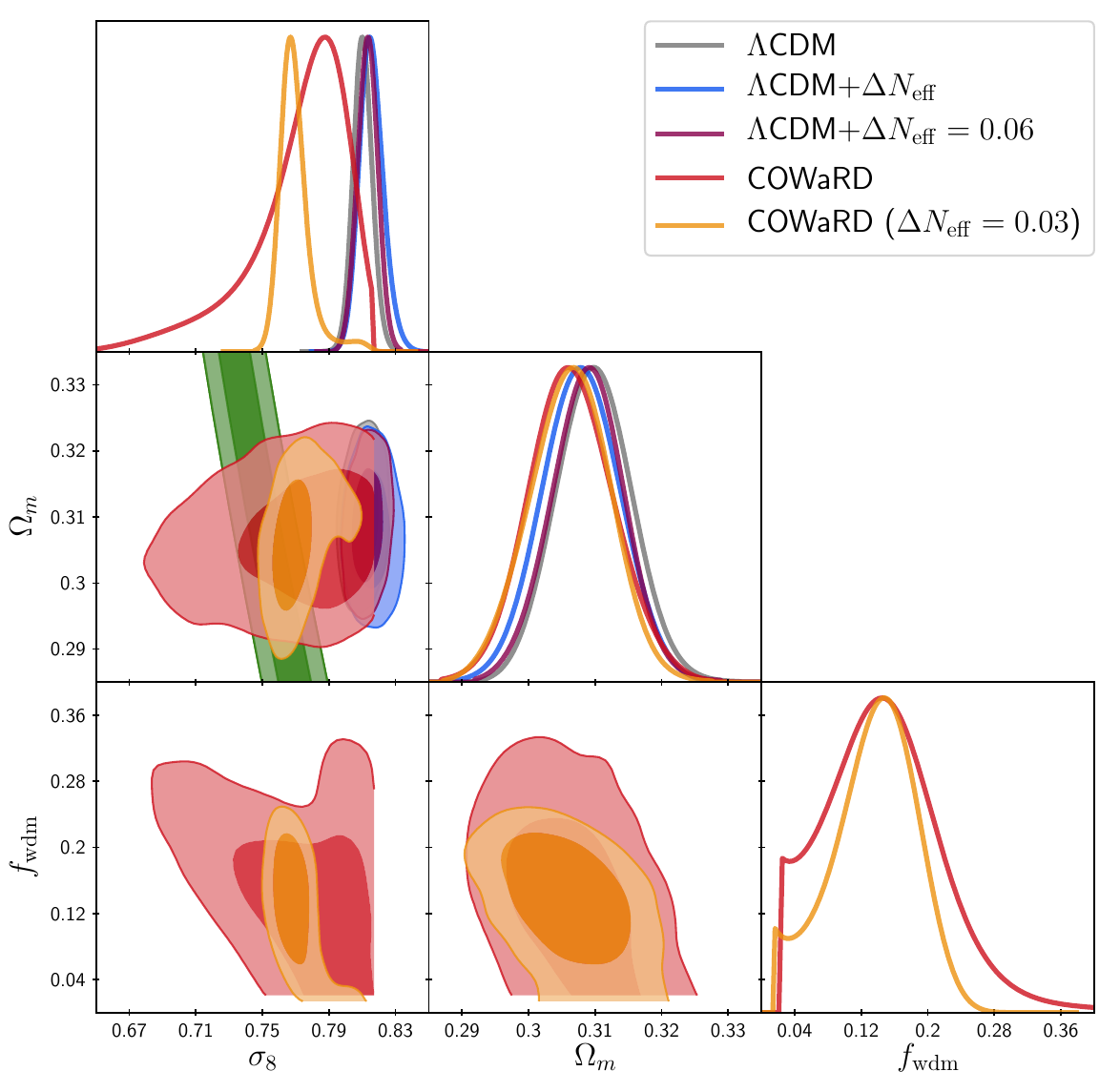}
\caption{Posteriors and likelihood contours of the $\sigma_8$, $\Omega_m$ and $f_\wdm$ parameters, from the fit of the \LC~(gray), COWaRD (red), and COWaRD with fixed $\DN=0.03$ (yellow) models to the cosmological data. For comparison, we also include the \LC+$\DN$ extension, with free $\DN$ (blue) and fixed $\DN=0.06$ (purple). In green are the $68\%$ and $95\%$ C.L. contours from the Planck SZ cluster counts, which measure $\sigma_8$ \cite{Ade:2015fva}.}
\label{fig:s8omf32anr}
\end{figure}

Perhaps the most important consequence of the COWaRD model is the suppression of LSS. \Fig{fig:s8omf32anr} shows the posteriors and likelihoods contours of the $\sigma_8$, $\Omega_m$ and $f_\wdm$ parameters resulting from the MCMC fits to the data. We consider the vanilla \LC~model and its extension \LC+$\DN$, as well as \LC+$\DN$ with fixed $\DN=0.06$, which corresponds to the COWaRD model with $\DN=0.03$ in its warm limit; we also consider the simplified COWaRD model (parameterized by $f_\wdm$ and $\DN$ and with $a_\dec=10^{-7}$), and the simplified COWaRD with fixed $\DN=0.03$. Also included for reference in the figure are the $68\%$ and $95\%$ C.L. contours from the Planck SZ cluster counts, which measure $\sigma_8 \bl( \Omega_m/0.27 \br)^{0.3} = 0.782 \pm 0.010$ \cite{Ade:2015fva}. As mentioned in \Sec{sec:tensions}, there is a mild tension between \LC~fits to Planck's CMB data and direct measurements of LSS, which is parameterized by $\sigma_8$ \cite{Aghanim:2018eyx,Heymans:2013fya,Hildebrandt:2016iqg,Joudaki:2016mvz,Troxel:2018qll,Troxel:2017xyo,Ade:2015fva,Ade:2013lmv,Kohlinger:2017sxk,Joudaki:2017zdt,MacCrann:2014wfa,Bohringer:2014ooa,Boehringer:2017wvr}, with Planck predicting a larger $\sigma_8$ than warranted by direct observations. As can be seen from the plot, the fit of the COWaRD model to the data automatically allows for smaller values of $\sigma_8$ {\it without having to include the measurements of LSS in the fit}, and, as shown in \Tab{tab:allchi2} in \App{appB}, without degrading the fit to the cosmological data when compared to \LC.

\begin{figure}
\centering
\includegraphics[width=0.6\textwidth]{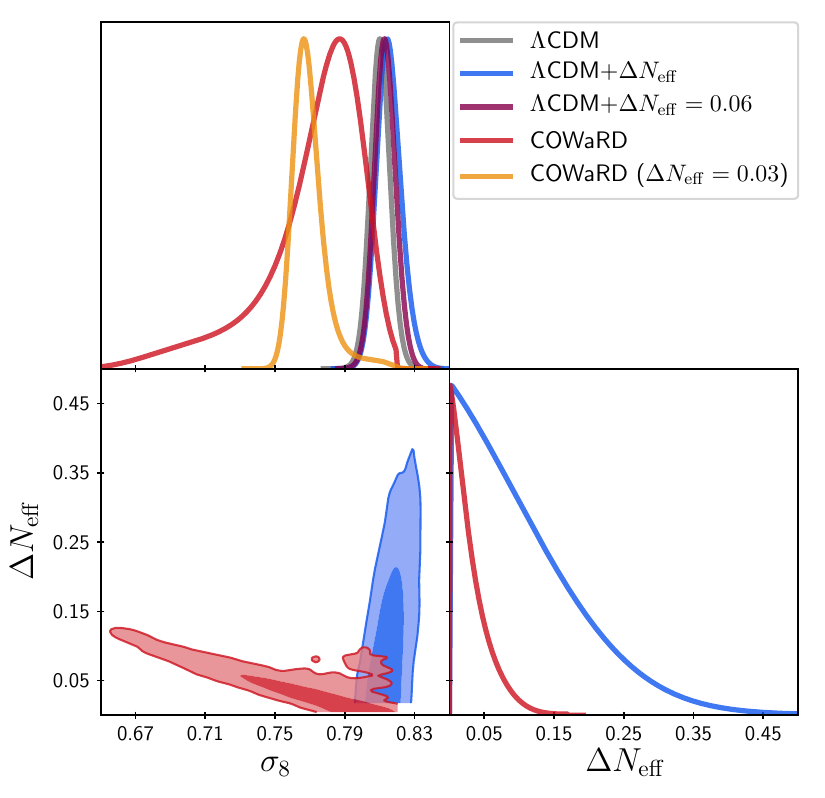}
\caption{Posteriors and likelihood contours of the $\sigma_8$ and $\DN$ parameters, from the fit of the \LC~(gray), COWaRD (red), and COWaRD with $\DN=0.03$ (yellow) models to the cosmological data. Also included are the \LC~extensions \LC+$\DN$ (blue) and \LC+$\DN=0.06$ (purple).}
\label{fig:s8dN}
\end{figure}

This brings us to the second consequence of the COWaRD model: $\DN$ and $\sigma_8$ are inversely correlated, as can be seen in \Fig{fig:s8dN}. This is in sharp contrast to what happens in the ``pure DR'' \LC$+\DN$ extension, also included in the figure, where $\sigma_8$ and $\DN$ are {\it positively} correlated \cite{Aghanim:2018eyx}. Note that this pure DR limit, when compared to \LC~there is only extra DR and no WDM, is not preferred by our fit to the data; as can be seen from \Tab{tab:allchi2}, the pure DR limit is disfavored by the high-$\ell$ TTTEEE data. This presents us with an exciting prospect: a positive detection of a non-vanishing $\DN$ immediately implies, within the COWaRD model, a value of $\sigma_8$ smaller than the one predicted by \LC; and vice versa. Future S4-CMB experiments will reach the threshold $\DN \approx 0.03$ \cite{Abazajian:2016yjj}, whereas next-generation LSS such as LSST and EUCLID \cite{Abell:2009aa,Laureijs:2011gra} surveys will measure $\sigma_8$ and $\Omega_m$ more precisely and decide the issue of the LSS tension. For example, were a value of $\DN=0.03$ to be discovered, the fit to the data of the COWaRD model boldly predicts $\sigma_8 = 0.770_{-0.010}^{+0.006}$ (see \Tab{tab:allpars} in \App{appB}). Alternatively, if a value of $\DN = 0.1$ were to be detected, the COWaRD fit would be degraded and our model disfavored, since it would predict $\sigma_8$ too small a value for LSS, e.g.~$\sigma_8 \sim 0.67$ too small even for the LSS tension.

COWaRD can be further probed by measurements of the matter power spectrum at smaller scales. The $1\sigma$ preferred values of $f_{\rm wdm}$ is above 0.05 for $\Delta N_{\rm eff}=0.03$. As can be seen in the left panel of Fig.~\ref{fig:Pk}, the matter power spectrum at $k \gg 1$ h/Mpc is suppressed more than 40\%. This large suppression can be detected by future measurements of 21-cm power spectrum~\cite{vanHaarlem:2013dsa, Koopmans:2015sua, DeBoer:2016tnn, Eastwood:2019rwh, Munoz:2019hjh}.

\section{Conclusions}\label{sec:concl}

In this work we have presented a cosmological model in which the dark radiation and warm dark matter, two of the most commonly studied extensions to the stanrdard $\LC$ model, have a common origin as the early decay products of an unstable massive particle. This simple scenario, which we dubbed COWaRD and is well motivated in particle physics, has effects on cosmological observables that are different from those effects that arise from the naive combination of dark radiation and warm dark matter.

Among these, the inversion of the well-known correlation between the amount of DR, parameterized by $\DN$, and the amount of large scale structure, parameterized by $\sigma_8$, is the most striking. Indeed, while in most dark radiation scenarios a positive $\DN$ indirectly results in an increase of matter abundance and thus of $\sigma_8$; our COWaRD model correlates the dark radiation abundance with the time at which the warm dark matter becomes non-relativistic. Thus, larger $\DN$ corresponds to larger free-streaming lenghts for the warm dark matter, which results in a suppression of large-scale structure and therefore a smaller $\sigma_8$. Future S4-CMB experiments will probe and maybe even detect small values of $\DN$, which would provide a way to falsify our COWaRD scenario: non-zero $\DN$ immeditaley requires, within our framework, an anti-correlated signal in $\sigma_8$. The fact that there is currently a tension between indirect and direct determinations of $\sigma_8$ makes this a tantalizing prospect.

Finally we want to draw attention to a conclusion of a more general nature. In cosmology it is common to consider extensions to the $\LC$ framework, such as dark radiation or warm dark matter, as stand-alone phenomenological models whose observable consequences are firmly established. However, as COWaRD demonstrates, the actual realization or embedding of these extensions into particle physics models can yield results that are different from, if not in direct contradiction with, those that could be drawn from the naive combination of said extensions. We therefore consider the overlap between cosmological and particle physics models as fertile ground for interesting and novel effects on cosmological observables.

{\bf Acknowledgments.}---%
We thank Lawrence Hall, Benjamin Horowiz, and Kathryn Zurek for collaboration in the early stages of this work, David Pinner for useful discussions, and Thejs Brinckmann for {\tt MontePython} technical support. Part of this research was conducted using computational resources and services at the Center for Computation and Visualization at Brown University. M.B.A. would like to thank  Boston University for its hospitality, and the Boston University Shared Computing Cluster (SCC) in the MGHPCC for the computing resources used in the early stages of this project. The work was supported in part by the NASA grant 80NSSC18K1010 (M.B.A.), the DoE Early Career Grant DE-SC0019225 (R.C.), the DoE grant DE-SC0009988 (K.H.) and the Raymond and Beverly Sackler Foundation Fund (K.H.).

\appendix

\section{Production of Axinos and Gravitinos}
\label{sec:prod_parent}
\renewcommand{\thesubsection}{A.\arabic{subsection}}

In this appendix, we review the computation of the relic abundance of the axino and the gravitino before decaying into WDM and DR. We assume that reheating after inflation completes at a temperature $T_{\rm RH}$, and the universe is radiation dominated until the standard matter-radiation equality.

\subsection{Production during Inflationary Reheating}
Axinos couple to gluons and gluinos by a dimension-5 coupling suppressed by the scale $ \sim 4\pi f_a / \alpha$ with $f_a$ the axion decay constant. Axinos are then dominantly produced around $T = T_{\rm RH}$. The number density of axinos normalized by the entropy density $s$ is~\cite{Strumia:2010aa}
\begin{align}
\left(\frac{n_{\tilde{a}}}{s}\right)_{\rm RH} \simeq 2\times 10^{-11}  \left( \frac{T_{\rm RH}}{10^4~{\rm GeV}} \right) \left( \frac{10^{13}~{\rm GeV}}{f_a}\right)^2 ,
\end{align}
which implies that the required reheat temperature is
\begin{equation}
\label{eq:TRH_axino}
T_{\rm RH} \simeq 20~{\rm TeV} \left( \frac{f_{3/2}}{0.1} \right) \left( \frac{\GeV}{m_{3/2}} \right)  \left( \frac{f_a}{10^{13}~{\rm GeV}}\right)^2 ,
\end{equation}
in order to generate the amount of gravitino WDM in units of total dark matter abundance, $f_{3/2} \equiv \Omega_{3/2}/ \Omega_{\rm DM}$.

Gravitinos couple to gluons and gluinos by a dimension-5 coupling suppressed by a scale $\sim m_{3/2} \mpl / m_{\tilde{g}}$ with $m_{\tilde{g}}$ the gluino mass. Gravitinos are also dominantly  produced around $T = T_{\rm RH}$, resulting in the number density~\cite{Moroi:1993mb, Rychkov:2007uq}
\begin{align}
\left(\frac{n_{3/2}}{s}\right)_{\rm RH} \simeq 4\times 10^{-9}  \left(\frac{25~\GeV}{m_{3/2}}\right)^2 \left( \frac{T_{\rm RH}}{10^{11}~{\rm GeV}} \right) \left( \frac{m_{\tilde{g}}}{{\rm TeV}}\right)^2.
\end{align}
To produce $f_{\tilde{a}} \equiv \Omega_{\tilde{a}}/ \Omega_{\rm DM}$ fraction of dark matter in warm axinos, the required reheat temperature is
\begin{equation}
\label{eq:TRH_gravitino}
T_{\rm RH} \simeq 3 \times10^8~\GeV \left( \frac{f_{\tilde{a}}}{0.1} \right) \left( \frac{4~\GeV}{m_{\tilde{a}}} \right) \left( \frac{m_{3/2}}{25~{\rm GeV}}\right)^2  \left( \frac{{\rm TeV}}{m_{\tilde{g}}}\right)^2 .
\end{equation}

If correlated signals in $\Delta N_{\rm eff}$ and $\sigma_8$ are detected, one can therefore infer the values of the reheat temperature after inflation by Eqs.~(\ref{eq:TRH_axino}) and (\ref{eq:TRH_gravitino}). The discussion in this section assumes radiation domination between the end of reheating until matter-radiation equality. The required reheat temperature can be much higher if there exists a matter-dominated epoch, which dilutes the relic abundance with entropy production. For example, Refs.~\cite{Banks:2002sd,Kawasaki:2008jc,Hasenkamp:2010if, Baer:2011eca, Bae:2014rfa, Co:2016fln, Co:2017orl} consider the case where a matter-dominated era originates from the condensate of the Pecci-Quinn symmetry breaking field.

\subsection{Production after the Freeze-out of the LOSP}
The lightest particle among the superpartners of the Standard Model particles other than the graviton is called the lightest observable supersymmetric particle (LOSP). After the LOSP abundance is fixed by the freeze-out process, it decays into the gravitino or the axino. This contribution can dominate over the above contribution if the LOSP annihilation is ineffective, which is the case for the bino-like LOSP. Assuming that the annihilation rate is dominated by t-channel exchange of nearly-degenerated right-handed sleptons $\tilde{\ell}_R$, the number density of the bino after the freeze-out is given by~\cite{ArkaniHamed:2006mb}
\begin{align}
\frac{n_{\tilde{B}}}{s} \simeq 2 \times 10^{-10} \left( \frac{m_{\tilde{\ell}_R}}{4~{\rm TeV}} \right)^4 \left( \frac{2~{\rm TeV}}{m_{\tilde{B}}} \right)^3 .
\end{align}
After the chain of decays $\tilde{B} \rightarrow \tilde{a}/\tilde{G} \rightarrow \tilde{G}/\tilde{a}$, this results in a fractional abundance of WDM in units of total dark matter abundance,
\begin{equation}
\frac{\Omega_{\rm wdm} h^2}{\Omega_{\rm DM} h^2} = 0.5 \left( \frac{m_{\rm wdm}}{\GeV} \right) \left( \frac{m_{\tilde{\ell}_R}}{4~{\rm TeV}} \right)^4 \left( \frac{2~{\rm TeV}}{m_{\tilde{B}}} \right)^3 ,
\end{equation}
with $m_{\rm wdm}$ the mass of warm dark matter, the lighter of $\tilde{a}$ and $\tilde{G}$.

The bino decay rate into the axino is~\cite{Covi:2001nw}
\begin{align}
\Gamma_{\tilde{B} \rightarrow \tilde{a} + B}  & = \frac{\alpha_{\mathrm{EM}}^2 C^2}{128 \pi^3 \cos ^{4} \theta_{W}} \frac{m_{\tilde{B}}^3}{f_{a}^2}\left(1-\frac{m_{\tilde{a}}^2}{m_{\tilde{B}}^2}\right)^3,
\end{align}
where the model-dependent parameter $C$ depends on the axion coupling with the hyper gauge bosons, the mass mixing of binos, and the Weinberg angle. The temperature at which the decay occurs in a radiation-dominated epoch is given by
\begin{align}
T_{\tilde{B} \rightarrow \tilde{a} + B} & \simeq 10~{\rm MeV} \ C \left( \frac{m_{\tilde{B}}}{\rm TeV} \right)^{{ \scalebox{1.01}{$\frac{3}{2}$} }} \left( \frac{10^{12}~{\rm GeV}}{f_a}\right) .
\end{align}
This implies that the gravitino becomes non-relativistic at the temperature
\begin{equation}
T_{\tilde{a}{\rm , nr}} \simeq 50~{\rm keV} \ C \left( \frac{m_{\tilde{B}}}{\rm TeV} \right)^{{ \scalebox{1.01}{$\frac{1}{2}$} }} \left( \frac{m_{\tilde{a}}}{6~\GeV} \right) \left( \frac{10^{12}~{\rm GeV}}{f_a}\right) .
\end{equation}
As assumed in deriving Eq.~(\ref{eq:warm_gravitino}), the axino is non-relativistic when decaying to the gravitino, $T_{\tilde{a}{\rm , nr}} > T_{\tilde{a} \rightarrow \tilde{G} a}$, as long as
\begin{equation}
f_a < 10^{16}~\GeV \ C \left( \frac{m_{\tilde{B}}}{\rm TeV} \right)^{{ \scalebox{1.01}{$\frac{1}{2}$} }} \left( \frac{m_{3/2}}{\rm GeV}\right)   \left( \frac{6~\GeV}{m_{\tilde{a}}} \right)^{{ \scalebox{1.01}{$\frac{3}{2}$} }} .
\end{equation}

Similarly, the decay rate and the temperature for the bino decay into the gravitino are~\cite{Covi:2001nw}
\begin{align}
\Gamma_{\tilde{B} \rightarrow \tilde{G} + Z/\gamma} & = \frac{m_{\tilde{B}}^5}{96\pi m_{3/2}^2 M_{\rm Pl}^2}  , \\
T_{\tilde{B} \rightarrow \tilde{G} + Z/\gamma} & \simeq 1~{\rm MeV} \left( \frac{m_{\tilde{B}}}{\rm TeV} \right)^{5/2} \left( \frac{\rm GeV}{m_{3/2}}\right) .
\end{align}
and the axino becomes non-relativistic at the temperature
\begin{equation}
T_{\rm 3/2, nr} \simeq {\rm keV} \left( \frac{m_{\tilde{B}}}{\rm TeV} \right)^{{ \scalebox{1.01}{$\frac{3}{2}$} }} ,
\end{equation}
which implies that, in deriving Eq.~(\ref{eq:warm_axino}), the assumption of a non-relativistic gravitino at the time of the decay, $T_{\rm 3/2, nr} > T_{\tilde{G} \rightarrow \tilde{a} a}$, is valid as long as $m_{\tilde{B}} > m_{3/2}$, which is anyway required by kinematics.

\section{MCMC Numerical Results}\label{appB}
\renewcommand{\thesubsection}{B.\arabic{subsection}}

In this appendix we summarize the numerical results from the MCMC fits of the \LC~and COWaRD models to the cosmological data. \Tab{tab:allchi2} presents the $\chi^2$ of the models, and \Tab{tab:allpars} shows the values of the different parameters.

\begin{table}[!ht]
	\begin{tabular}{|| c || c | c | c | c | c ||}
		\hline
		\multicolumn{6}{|c|}{Best-fit $\chi^2$} \\
		\hline\hline
    Data Sets & \LC & \LC+$\DN$ & \LC+$\DN$ \tiny{($=0.06$)} & COWaRD & COWaRD \tiny{($\DN=0.03$)} \\ [0.5ex]
		\hline\hline
		high-$\ell$ TTTEEE & $2347.56$ & $2348.03$ & $2349.39$ & $2345.94$ & $2346.38$ \\
		\hline
    low-$\ell$ EE & $396.37$ & $396.72$ & $396.35$ & $395.95$ & $395.96$ \\
		\hline
    low-$\ell$ TT & $22.86$ & $22.96$ & $22.27$ & $22.47$ & $22.25$ \\
    \hline
    lensing & $8.87$ & $8.79$ & $9.06$ & $10.11$ & $10.45$ \\
    \hline
    Pantheon & $1027.49$ & $1027.08$ & $1027.05$ & $1027.06$ & $1027.02$ \\
		\hline
		BAO & $5.57$ & $5.35$ & $5.32$ & $5.29$ & $5.24$ \\
		\hline\hline
		TOTAL & $3808.72$ & $3808.92$ & $3809.44$ & $3806.82$ & $3807.28$ \\
		\hline
		$\Delta \chi^2_\mathrm{eff}$ & --- & $+0.2$ & $+0.72$ & $-1.9$ & $-1.44$ \\ [1ex]
		\hline\hline
	\end{tabular}
	\caption{Minimum {\it effective chi square} $\chi^2_\mathrm{eff} =-2\ln {\cal L}$ of the \LC~and COWaRD models.\label{tab:allchi2}}
\end{table}

\begin{table}[!ht]
	\begin{tabular}{|| c || c | c | c | c | c ||}
		\hline
		\multicolumn{6}{|c|}{Parameter values} \\
		\hline\hline
    Parameter & \LC & \LC+$\DN$ & \LC+$\DN$ \tiny{($=0.06$)} & COWaRD & COWaRD \tiny{($\DN=0.03$)} \\ [0.5ex]
		\hline\hline
		$100~\omega_{b }$ & $2.245_{-0.014}^{+0.014}$ & $2.252_{-0.015}^{+0.015}$ & $2.249_{-0.014}^{+0.013}$ & $2.254_{-0.015}^{+0.015}$ & $2.254_{-0.015}^{+0.014}$ \\
		\hline
    $n_{s }$ & $0.968_{-0.004}^{+0.004}$ & $0.971_{-0.005}^{+0.004}$ & $0.970_{-0.004}^{+0.004}$ & $0.970_{-0.004}^{+0.004}$ & $0.970_{-0.004}^{+0.004}$ \\
		\hline
    $\tau_\mathrm{reio}$ & $0.056_{-0.008}^{+0.007}$ & $0.057_{-0.008}^{+0.007}$ & $0.057_{-0.007}^{+0.007}$ & $0.052_{-0.008}^{+0.008}$ & $0.052_{-0.009}^{+0.008}$ \\
    \hline
    $100~\theta_s$ & $1.0420_{-0.00029}^{+0.00029}$ & $1.0417_{-0.00033}^{+0.00037}$ & $1.0418_{-0.00029}^{+0.00028}$ & $1.0419_{-0.00031}^{+0.00032}$ & $1.0419_{-0.00030}^{+0.00030}$ \\
    \hline
    $\ln 10^{10}A_s$ & $3.048_{-0.015}^{+0.014}$ & $3.053_{-0.016}^{+0.014}$ & $3.051_{-0.014}^{+0.015}$ & $3.037_{-0.017}^{+0.016}$ & $3.037_{-0.018}^{+0.017}$ \\
		\hline
		$\omega_\cdm$ & $0.1192_{-0.0009}^{+0.0009}$ & $0.1210_{-0.0020}^{+0.0012}$ & $0.1202_{-0.0010}^{+0.0009}$ & $0.1029_{-0.0097}^{+0.0096}$ & $0.1038_{-0.0087}^{+0.0056}$ \\
    \hline
    $\Delta N_\mathrm{eff}$ & --- & $0.110_{-0.110}^{+0.028}$ & 0.06 & $0.031_{-0.031}^{+0.007}$ & 0.03 \\
    \hline
    $f_\wdm$ & --- & --- & --- & $0.138_{-0.082}^{+0.074}$ & $0.130_{-0.044}^{+0.067}$ \\
		\hline\hline
    $H_0$ & $67.77_{-0.43}^{+0.43}$ & $68.43_{-0.77}^{+0.55}$ & $68.13_{-0.41}^{+0.42}$ & $68.19_{-0.53}^{+0.47}$ & $68.21_{-0.48}^{+0.46}$ \\
    \hline
    $\sigma_8$ & $0.801_{-0.006}^{+0.006}$ & $0.816_{-0.008}^{+0.007}$ & $0.813_{-0.006}^{+0.006}$ & $0.773_{-0.014}^{+0.038}$ & $0.770_{-0.010}^{+0.006}$ \\
    \hline
    $\Omega_m$ & $0.310_{-0.006}^{+0.006}$ & $0.308_{-0.006}^{+0.006}$ & $0.309_{-0.006}^{+0.005}$ & $0.307_{-0.007}^{+0.006}$ & $0.306_{-0.006}^{+0.006}$ \\
    \hline\hline
	\end{tabular}
	\caption{Mean values and 68\% C.L. intervals of the parameters of the \LC~and COWaRD models.\label{tab:allpars}}
\end{table}

\newpage

\bibliography{coward}{}

\end{document}